\documentclass[twocolumn,trackchanges, twocolappendix]{aastex7}

\usepackage{amsmath}
\usepackage{physics}
\usepackage{bm}
\usepackage{multirow}

\usepackage{color}
\hypersetup{colorlinks=true, linkcolor=black, citecolor=blue,}

\begin{document}

\title{Proposal of a Novel Physical Parameter Characterizing Solar Wind Speed in a Wave-Driven Model}

\author[orcid=0009-0007-8101-9668, gname=Kyogo, sname=Tokoro]{Kyogo Tokoro} 
\affiliation{The University of Tokyo, Department of Earth and Planetary Science}
\email{t.kyogo1056.0706@eps.s.u-tokyo.ac.jp}

\author[orcid=0000-0002-7136-8190, gname=Munehito, sname=Shoda]{Munehito Shoda} 
\affiliation{The University of Tokyo, Department of Earth and Planetary Science}
\email{shoda.m.astroph@gmail.com}

\author[orcid=0000-0001-7891-3916, gname=Shinsuke, sname=Imada]{Shinsuke Imada} 
\affiliation{The University of Tokyo, Department of Earth and Planetary Science}
\email{imada@eps.s.u-tokyo.ac.jp}

\begin{abstract}

Empirical solar wind speed models play an important role in enabling space weather forecasting with low computational cost. Among these, one model called WS model is based on the asymptotic expansion factor. However, it is known that it fails in the case of pseudostreamers. In this study, as a first step toward constructing a solar wind speed empirical model based on physical parameters, we investigated the effect of the radial profile of flux-tube shape on the solar wind speed using one-dimensional numerical simulations. In the simulations, ad hoc Alfvén waves are injected from the photosphere at $r=R_\odot$ as the energy source, and the MHD equations are solved out to the interplanetary space at $r=70R_\odot$ to reproduce solar wind acceleration. As a result, even when the coronal base magnetic field and the asymptotic expansion factor are fixed, the final solar wind speed varies by approximately $300$ km s$^{-1}$ depending on changes in the expansion height or non-monotonic expansion. Additionally, across all simulations performed, a better correlation is found with the quantities that reflect the information about the radial profile of flux-tube shape than the asymptotic expansion factor. Our results suggest that, as a physical characteristic parameter of the solar wind speed, an operation that can account for the expansion factor throughout the corona is necessary.

\end{abstract}

\section{Introduction} \label{sec:intro}


The solar wind \citep{Parker_1958_ApJ, Neugebauer_1962_Sci, Velli_1994_ApJ} is one of the key topics in solar and heliospheric physics. In particular, in the heliosphere, it causes geomagnetic disturbances \citep{Ebihara_2006_JGR}, posing a potentially serious threat to modern civilization expanding into space. This potential impact on human society has led to a growing recognition of the importance of space weather forecasting \citep{Schwenn_2006_LRSP, Pulkkinen_2007_LRSP, Temmer_2021_LRSP}, which aims to predict disturbances in planetary magnetospheres and ionospheres caused by the solar wind.


In space weather forecasting, the spatiotemporal structures of velocity, density, temperature and magnetic field of the solar wind are important parameters \citep{Arge_2010_AIPCS,Ebihara_2006_JGR}. In particular, wind velocity causes significant events. Fast solar wind from a large coronal hole forms a co-rotating interaction region (CIR) with the preceding slow wind \citep{Smith_1976_GRL,Gosling_1978_JGR, Richardson_2018_LRSP}, frequently causing moderate to small-scale geomagnetic disturbances \citep{Richardson_2012_JSWSC}. It also interacts with Coronal Mass Ejections (CMEs), accelerating or decelerating them depending on the relative velocity, making it a crucial factor in predicting large-scale geomagnetic storms \citep{Gosling_1991_JGR,Richardson_2012_JSWSC,Shiota_2016_SW,Wold_2018_JSWSC}.


Since magnetic field is supposed to be a key factor in characterizing the solar wind properties, many studies have focused on magnetic fields \citep{Wang_1990_ApJ, Arge_2003_AIP}. In space weather forecasting, it is necessary to predict the solar wind speed with sufficient lead time. Therefore, practical prediction methods are limited to those that can be executed with low computational cost. To this end, empirical models derived from observations or simplified simulations are commonly used in the model-based space weather prediction. One example of the wind speed prediction models is called the Wang–Sheeley (WS) model \citep{Wang_1990_ApJ, Arge_2000_JGR}. Solar wind speed of WS model is determined only by asymptotic the expansion factor of magnetic flux tubes, $f_\infty$, which characterizes the extent to which the cross-sectional area of the tube expands from the coronal base to interplanetary space, and this model is used as the inner boundary condition for the solar wind speed in several heliosphere models \citep{Odstrcil_2003_ASR, Shiota_2014_SW, Shiota_2016_SW}. 

It has been pointed out, however, that the WS model overestimates poorly the solar wind speed from pseudostreamers \citep{Wang_2007_ApJ, Riley_2012_SoPh, Wang_2012_ApJ, Riley_2015_SpW}. Pseudostreamers are the regions between the coronal holes with the same magnetic polarity and have non-monotonic flux-tube shape. Simply looking at $f_\infty$ does not capture the effects of structures such as pseudostreamers, which exhibit relatively small $f_\infty$ values. Another empirical model is the Wang–Sheeley–Arge (WSA) model \citep{Arge_2003_AIP, Arge_2004_JASTP, McGregor_2011_JGRA}, which adds the distance from the coronal hole boundary \citep[DCHB,][]{Riley_2001_JGR} as an additional characteristic parameter. The WSA model extends the WS model by including the idea that the solar wind speed decreases with smaller value of DCHB, and it is often used as the inner boundary condition in the modern space weather models \citep{Cohen_2007_ApJ, Pahud_2012_JASTP, Mays_2015_SoPh, Pomoell_2018_JSWSC, Shen_2018_ApJ, Verbeke_2022_AandA}. Although the WSA model is considered to reasonably reproduce the solar wind speed, it involves numerous free parameters whose optimal values vary significantly from one Carrington Rotation to another \citep{Riley_2015_SpW}. Moreover, the DCHB parameter inherent in the WSA model is highly sensitive to small-scale magnetic features such as small loops within coronal holes \citep{Wang_2020_ApJ}, causing the optimized parameters to differ substantially depending on the resolution of the magnetogram \citep{Majumda_2025_arXiv}.

In recent years, in addition to empirical models, wave/turbulence-driven models have been increasingly used to physically model the global structure of the solar wind \citep{Usmanov_2018_ApJ, van_der_Holst_2019_ApJ, Reville_2020_ApJS, Reville_2022_AandA}, achieving considerable success. Among them, the AWSoM model \citep{Sokolov_2013_ApJ, van_der_Holst_2014_ApJ, Meng_2015_MNRAS}, a widely used standard, has been optimized for real-time simulations to support space weather forecasting \citep{Sokolov_2021_ApJ, Liu_2025_ApJ}. However, these global models employ mean-field approximations based on reduced MHD, thereby neglecting compressive waves that possibly plays a crucial role in solar wind acceleration \citep{Suzuki_2005_ApJ, Matsumoto_2014_MNRAS, Shoda_2019_ApJ}. In addition, global models require the specification of the Poynting flux and turbulent correlation length at the inner boundary, which remain difficult to measure \citep[but see][]{Sharma_2023_NatAs, Bailey_2025_ApJ, Morton_2025_ApJ} but significantly influence the simulation results \citep{Jivani_2023_SpWea}.

The flux-tube model, which focuses on a single open flux tube, is often used as an alternative to global models. It is particularly effective for understanding physical processes occurring within the solar wind \citep{Perez_2013_ApJ, van_Ballegooijen_2016_ApJ, van_Ballegooijen_2017_ApJ, Shoda_2018_ApJ_frequency_dependent, Chandran_2019_JPlPh}. Additionally, it enables a self-consistent modeling from the photosphere to the solar wind, incorporating compressibility \citep{Suzuki_2005_ApJ}. In terms of wind speed modeling, it offers flexibility in controlling the expansion factor, a key parameter influencing wind speed, making it a powerful tool for investigating their relationship \citep{Wang_1994_ApJ, Shoda_2022_ApJ}. This advantage is crucial for structures such as pseudostreamers, where the flux-tube expansion is known to be non-monotonic \citep{Panasenco_2013_AIPC, Panasenco_2019_ApJ}.

In this study, we investigate how the expansion factor as a function of $r$, $f(r)$, affects the solar wind speed using high-resolution numerical simulations sufficient to resolve Alfvén waves. To directly examine the physical influence of $f(r)$ on the solar wind speed, a one-dimensional calculation is necessary so that the artificially imposed $f(r)$ does not change over time. We also investigate which parameters related to $f(r)$ may serve as characteristic quantities that strongly correlate with the solar wind speed.


\section{Method}

\subsection{Basic Equations}
In this study, we investigate a one-dimensional system along a thin, static open magnetic flux tube. Magnetohydrodynamic simulations in such systems are frequently employed in numerical studies of coronal heating \citep{Moriyasu_2004_ApJ, Washinoue_2019_ApJ, Shoda_2021_AandA, Washinoue_2021_MNRAS, Matsumoto_2024_ApJ} and solar wind acceleration \citep{Suzuki_2005_ApJ, Suzuki_2006_JGR, Chandran_2011_ApJ, Shoda_2018_ApJ, Shoda_2020_ApJ, Shoda_2023_ApJ, Matsuoka_2024_ApJ}. Assuming that the magnetic flux tube extends radially, we use the solar radial distance ($r$) as the coordinate along the flux tube. Under these assumptions, the governing equations are given as follows \citep[see Appendix A of][for derivation]{Shoda_2021_AandA}.
\begin{align}
    \frac{\partial}{\partial t} \rho + \frac{1}{A} \frac{\partial}{\partial r} \left( \rho v_r A \right) = 0, \label{eq:mass_conservation}
\end{align}
\begin{align}
    \frac{\partial}{\partial t} \left( \rho v_r \right) &+ \frac{1}{A} \frac{\partial}{\partial r} \left[ \left( \rho v_r^2 + p + \frac{\boldsymbol{B}_\perp^2}{8\pi}\right) A \right] \nonumber \\
    &= \left( \frac{1}{2} \rho \boldsymbol{v}_\perp^2 + p \right) \frac{d}{dr} \ln A - \rho \frac{GM_\odot}{r^2},
    \label{eq:r momentum}
\end{align}
\begin{align}
    \frac{\partial}{\partial t} \left( \rho v_x \right) &+ \frac{1}{A} \frac{\partial}{\partial r} \left[ \left( \rho v_r v_x - \frac{B_r B_x}{4 \pi} \right) A \right] \nonumber \\ &= - \frac{1}{2} \left( \rho v_r v_x - \frac{B_r B_x}{4 \pi} \right) \frac{d}{dr} \ln A + \rho D_x^v,
\end{align}
\begin{align}
    \frac{\partial}{\partial t} \left( \rho v_y \right) &+ \frac{1}{A} \frac{\partial}{\partial r} \left[ \left( \rho v_r v_y - \frac{B_r B_y}{4 \pi} \right) A \right] \nonumber \\ &= - \frac{1}{2} \left( \rho v_r v_y - \frac{B_r B_y}{4 \pi} \right) \frac{d}{dr} \ln A + \rho D_y^v, 
\end{align}
\begin{align}
    \frac{\partial}{\partial t} B_x &+ \frac{1}{A} \frac{\partial}{\partial r} \Big[ \left( v_r B_x - B_r v_x \right) A \Big] \nonumber \\ &= \frac{1}{2} \left( v_r B_x - B_r v_x \right) \frac{d}{dr} \ln A + \sqrt{4 \pi \rho} D_x^b,
\end{align}
\begin{align}
    \frac{\partial}{\partial t} B_y &+ \frac{1}{A} \frac{\partial}{\partial r} \Big[ \left( v_r B_y - B_r v_y \right) A \Big] \nonumber \\ &= \frac{1}{2} \left( v_r B_y - B_r v_y \right) \frac{d}{dr} \ln A + \sqrt{4 \pi \rho} D_y^b, 
\end{align}
\begin{align}
    \frac{\partial}{\partial t} e &+ \frac{1}{A} \frac{\partial}{\partial r} \Bigg[ \bigg[ \left( e + p + \frac{\boldsymbol{B}_\perp^2}{8 \pi} \right) v_r - B_r \frac{\boldsymbol{v}_\perp \cdot \boldsymbol{B}_\perp}{4 \pi} \bigg] A \Bigg] \nonumber \\ &= - \rho v_r \frac{GM_\odot}{r^2} + Q_{\rm cnd} + Q_{\rm rad}, \label{eq:energy_conservation}
\end{align}
where $A$ denotes the cross section of the flux tube  and the notation of the other symbols follows that of \citet{Shoda_2021_AandA}.

Turbulent dissipation, which is thought to play a critical role in coronal holes and the solar wind, is not directly included in the one-dimensional system. Therefore, we introduce this effect phenomenologically \citep{Hossain_1995_PhFl, Dmitruk_2002_ApJ} through $D^v_{x,y}$ and $D^b_{x,y}$ given as below \citep{Shoda_2018_ApJ}.
\begin{equation}
    D^v_{x,y} = - \frac{c_d}{4\lambda_\perp} \left( z_\perp^+ z_{x,y}^- +  z_\perp^-  z_{x,y}^+  \right), \label{eq:phenomenological_awt_vsource}
\end{equation}
\begin{equation}
    D^b_{x,y} = - \frac{c_d}{4\lambda_\perp} \left( z_\perp^+ z_{x,y}^- - z_\perp^- z_{x,y}^+  \right), \label{eq:phenomenological_awt_bsource}
\end{equation}
\begin{equation}
    \lambda_\perp = \lambda_{\perp,\ast} \frac{r}{R_\odot} \sqrt{\frac{f}{f_\ast}}, \label{eq:lambda_perp_via_f}
\end{equation}
where
\begin{equation}
    z_{x,y}^\pm = v_{x,y} \mp B_{x,y}/\sqrt{4 \pi \rho}, \label{eq:elsasser_variables_definition}
\end{equation}
\begin{equation}
    z_\perp^\pm = \sqrt{{z_x^\pm}^2+{z_y^\pm}^2}
\end{equation}
are the Els\"asser variables \citep{Elsasser_1950_PhRv, Marsch_1987_JGR, Magyar_2019_ApJ} and $\lambda_\perp$ denotes the perpendicular correlation length \citep{Abramenko_2013_ApJ, Sharma_2023_NatAs}. Here, we assume that the perpendicular correlation length is identical for both the forward and backward Els\"asser variables and scales with the radius of the flux-tube cross section \citep[$\propto r\sqrt{f}$,][]{Hollweg_1986_JGR, Cranmer_2005_ApJS}, although we must acknowledge that this assumption may be incorrect \citep{Adhikari_2017_ApJ, Bandyopadhyay_2020_ApJS, Adhikari_2022_ApJ, Wu_2022_ApJ}. The non-dimensional parameter $c_d$ represents the efficiency of the energy cascade and is set to $c_d = 0.1$ based on theoretical \citep{Usmanov_2014_ApJ, van_Ballegooijen_2016_ApJ, Verdini_2019_SoPh} and observational \citep{Bandyopadhyay_2020_ApJS, Wu_2022_ApJ, Bowen_2024_arXiv} studies. We note that the heating rate resulting from these phenomenological terms is described as
\begin{align}
    Q_{\rm turb} = \rho c_d \frac{\left| z_\perp^- \right| {z_\perp^+}^2  + \left| z_\perp^+ \right| {z_\perp^-}^2 }{4 \lambda_\perp}.
\end{align}

$Q_{\rm cnd}$ represents heating by thermal conduction, expressed via the heat flux $q_{\rm cnd}$ as follows.
\begin{align}
    Q_{\rm cnd} = - \frac{1}{r^2 f} \frac{\partial}{\partial r} \left( q_{\rm cnd} r^2 f \right).
\end{align}
In regions such as the transition region and corona, where density is sufficiently high and particle collisions occur frequently, the Spitzer-Harm formulation \citep{Spitzer_1953_PhRv} is applicable for the heat flux. Meanwhile, in systems like the solar wind, where the mean free path of particles is comparable to the system size, the Spitzer-Harm flux is inapplicable \citep{Salem_2003_ApJ, Bale_2013_ApJ}, and the free-streaming flux has been proposed as an alternative \citep{Hollweg_1976_JGR, Cranmer_2021_JGR}. We therefore formulate the heat flux with a continuous transition from the Spitzer-Harm flux to the free-streaming flux as follows.
\begin{align}
    q_{\rm cnd} = \xi_{\rm cnd} q_{\rm SH} + \left( 1-\xi_{\rm cnd} \right) q_{\rm FS},
\end{align}
where $q_{\rm SH}$ and $q_{\rm FS}$ denote the Spitzer-Harm and free-streaming heat fluxes, respectively, given as follows.
\begin{align}
    q_{\rm SH} = - \kappa_{\rm SH} T^{5/2} \frac{\partial T} {\partial r}, \ \ \ \ q_{\rm FS} = \frac{3}{4} \alpha_{\rm FS} p v_r,
\end{align}
where $\kappa_{\rm SH} = 10^{-6}$ in cgs and we set $\alpha_{\rm FS} = 2$ following \citet{Cranmer_2021_JGR}. The bridging parameter $\xi_{\rm cnd}$ is employed as follows.
\begin{align}
    \xi_{\rm cnd} = \min \left[ 1, \left( \frac{r_{\rm cnd}}{r} \right)^2 \right],
\end{align}
where $q_{\rm SH}$ and $q_{\rm FS}$ represent the Spitzer-Harm heat flux and free-streaming heat flux, respectively. $r_{\rm cnd}$ represents the radius at which the heat flux transitions from Spitzer-Harm to free-streaming type. We set $r_{\rm cnd}/R_\odot = 5$.


$Q_{\rm rad}$ is the radiative loss function, consisting of optically thick ($Q_{\rm rad, thck}$) and optically thin ($Q_{\rm rad, thin}$) components. Ideally, the optically thick radiation should be solved using the (non-LTE) radiative transfer equation \citep{Gudiksen_2011_AandA, Iijima_2017_ApJ}; however, this approach is computationally expensive. Therefore, following previous studies \citep{Gudiksen_2005_ApJ, Isobe_2008_ApJ, Shoda_2021_AandA}, we approximate the optically thick cooling with an exponential cooling term, while the optically thin cooling is expressed as $Q_{\rm rad, thin} = n_{\rm H} n_e \Lambda(T)$, where $\Lambda (T)$ is the radiative loss function. For the detailed formulation, refer to Section 2.5 of \citet{Shoda_2023_ApJ}.

\subsection{Radial Profile of Expansion Factor}

Open flux tubes typically undergo the two-step super-radial expansion. The first stage arises in the chromosphere, where a decline in gas pressure due to stratification leads to the expansion of magnetic fields concentrated by convective collapse at the photosphere \citep{Keller_2004_ApJ, Cranmer_2005_ApJS, Ishikawa_2021_SciA}. The second stage is driven by the coronal magnetic field configuration: open tubes extend over closed loops, eventually reaching a filling factor of 1 \citep[see Figure 3 in][]{Cranmer_2011_ApJ}.

To describe the two-step expansion in terms of the cross section of the flux tube $A(r)$, we follow the formulation employed in \citet{Kopp_1976_SoPh} and \citet{Shoda_2023_ApJ} with a modification to account for the non-monotonic expansion:
\begin{align}
    A(r)&= \frac{A(R_\odot)r^2}{{R_\odot}^2}\eta(r)f(r), \label{eq:A} \\
    \eta(r)&=\frac{\eta_1(r)}{\sqrt{\eta_1(r)^2+(\eta_{\rm{exp}}^2-1)f(r)^2}}\eta_{\rm{exp}}, \\
    f(r)&=f_{\rm{KH}} (r) g(r), \label{eq:f}
\end{align}
where
\begin{align}
    \eta_1(r)&=
    \exp\qty(\frac{r-R_\odot}{2h_{\rm{exp}}} ), \\
    f_{\rm{KH}}(r)&=\frac{\{\mathcal{F}(r)-\mathcal{F}(R_\odot)\}f_{\infty}+\mathcal{F}(R_\odot)+1}{\mathcal{F}(r)+1} ,\\
    \mathcal{F}(r)&=\exp\qty(\frac{r-r_{\rm{exp}}}{\sigma_{\rm{exp}}}), \\
    g(r)&=
    1+(g_{\rm{max}}-1)\exp\qty(-\qty[\frac{\log \tilde {r_{\rm{g}}}(r)}{\log w}]^2), \label{eq:g(r)} \\
    \tilde {r_{\rm{g}}}(r)&=\frac{r/R_\odot-1+\epsilon}{r_{\rm{max}}/R_\odot-1} \quad (\epsilon \rightarrow +0). \label{eq:rg}
\end{align}
The first expansion in the chromosphere ($1 < r/R_\odot < 1.003$) is represented by the function $\eta(r)$, which causes the magnetic field to expand exponentially with a scale height of $2h_{\rm{exp}}$. Here, $h_{\rm{exp}}$ denotes the pressure scale height at the photosphere. This ensures that the plasma beta remains nearly constant throughout the expansion, up to the final cross-sectional expansion factor of $\eta_{\rm{exp}}$. The uncertainty introduced by this assumption has been discussed in the Appendix of \citet{Shoda_2022_ApJ} and can lead to differences in the velocity on the order of several tens of km s$^{-1}$. The expansion in the corona ($1.01 < r/R_\odot$) is represented by $f(r)$, with its final expansion factor represented by $f_\infty$. The parameters $r_{\rm{exp}}$ and $\sigma_{\rm{exp}}$ define the height and width of the coronal expansion, respectively. 

$g(r)$ represents the magnitude of the non-monotonicity in the flux-tube expansion. For monotonic expansion in the corona, $g_{\rm{max}} = 1$ and hence $g(r) \equiv 1$. When $g_{\rm{max}} > 1$, $g(r)$ represents a Gaussian-shape non-montonic expansion on the logarithmic scale of $r/R_\odot - 1$. The parameters $g_{\rm{max}}$, $r_{\rm{max}}$, and $w$ quantify the magnitude, central height, and thickness of the non-monotonic expansion, respectively. Here, $\epsilon$ is an infinitesimal positive value to avoid $\log 0$ in Equation \eqref{eq:g(r)}. The functions are normalized so that $\eta(R_\odot) = f(R_\odot) = 1$. Furthermore, it should be noted that the choice of $A(R_\odot)$ is arbitrary.

We also set the background magnetic field at the photosphere, $B_{r,\odot}$, as the boundary condition for the magnetic field, adopting $1340$ G as a reference value. Because of the magnetic flux conservation law, $B_r(r)$ is expressed as follows.
\begin{align}
    B_r(r) = \frac{B_{r,\odot}{R_\odot}^2}{\eta(r) f(r)r^2}.
\end{align}
The uncertainty arising from fixing $B_{r,\odot}$ to $1340$ G is discussed in Appendix \ref{app:B_rodot}. 

The free parameters varied in this study are summarized in Table~\ref{tab:free parameters}. To prevent the interplanetary magnetic field from becoming excessively weak, we only conducted simulations satisfying the condition $\eta_{\rm exp} f_{\infty} \ge 10^5$, equivalent to $B_{r,{\rm E}} > 0.028$ nT, where $B_{r,{\rm E}}$ denotes the (unsigned) radial magnetic field strength at 1 au.

\begin{deluxetable*}{c c c c}
\tablecaption{List of the free parameters in this study\label{tab:free parameters}}
\tablehead{
\colhead{} &
\colhead{Meaning} &
\colhead{Fiducial value} &
\colhead{Value range} 
}
\startdata
$B_{r,\odot}$ & radial magnetic field strength at photosphere & $1340$ G & ($167.5$–$1340$) G \\[3pt]
$\eta_{\rm exp}$ & asymptotic expansion factor in the chromosphere & $10^{2}$ & $10^{1}$–$10^{3}$ \\[3pt]
$f_\infty$   & asymptotic expansion factor in the corona  & $10^{1}$ & $10^{0}$–$10^{3}$  \\[3pt]
$r_{\rm exp}$ & height at which coronal expansion begins  & $1.3\,R_\odot$ & $1.05$–$3.0\,R_\odot$  \\[3pt]
$\sigma_{\rm exp}$ & radial scale of coronal expansion  & $0.5\,R_\odot$ & $0.1$–$1.0\,R_\odot$ \\[3pt]
$g_{\rm max}$ & degree of coronal non-monotonic expansion  & 1 & 1–16 \\[3pt]
$r_{\rm max}$ & height of the peak of coronal non-monotonic expansion & $1.25\,R_\odot$ & $1.125$–$2.0\,R_\odot$ \\[3pt]
$w$ & normalized scale of coronal non-monotonic expansion region\,(in $\log(r/R_\odot-1)$) & 2 & 2–8                  
\enddata
\end{deluxetable*}

\subsection{Boundary Condition}
The outer boundary of the simulation domain, typically at $r=70R_\odot$, is set beyond the Alfv\'en point. Since the thermal conduction near the outer boundary is dominated by the free streaming component, all physical information propagates outward there. Thus, the choice of the outer boundary condition has minimal impact on the simulation results. In this study, we impose the following boundary conditions for the stability of simulation.
\begin{align}
    \left. \frac{\partial}{\partial r} \left( \rho r^2 \right) \right|_{\rm out} = 0, \\
    \left. \frac{\partial}{\partial r} \left( pr^3 \right) \right|_{\rm out} = 0, \\
    \left. \frac{\partial}{\partial r} \boldsymbol{v} \right|_{\rm out} = 0, \\
    \left. \frac{\partial}{\partial r} \left(  \boldsymbol{B}_\perp r \right) \right|_{\rm out} = 0,
\end{align}
where $X_{\rm out}$ denotes the value of $X$ at the outer boundary. We also set an upper limit for $T_{\rm out}$ as $T_{\rm out} \le T_{\rm out}^{\rm UL} = 2 \times 10^{6} {\rm \ K}$ to ensure the numerical stability. We confirmed that in the quasi-steady state, $T_{\rm out}$ remains below $T_{\rm out}^{\rm UL}$ at all times, and hence, this upper limit does not affect the physical properties of the simulation results. 

The inner boundary condition was implemented as described in Section 2.7 of \citet{Shoda_2023_ApJ}. To account for sufficient mass-loss rates without flux emergence, the injected Poynting flux at the lower boundary was enhanced by roughly a factor of three ($F^{\rm AW}_{\rm in} = 1.6 \times 10^9 {\rm \ erg \ cm^{-2} \ s^{-1}}$) compared with \citet{Shoda_2023_ApJ}. The Alfv\'en-wave energy flux is defined as
\begin{align}
    F_{A}=\qty(\frac{1}{2}\rho \bm{v}_\perp^2+\frac{\bm{B}_\perp^2}{4\pi})v_r-\frac{B_r}{4\pi}\qty(\bm{v}_\perp\cdot \bm{B}_\perp), \label{eq:def F_A}
\end{align} 
and the net injected value $F^{\rm AW}_{\rm in}$ corresponds to this flux evaluated at the inner boundary. Across all simulations, the amplitudes of the transverse velocity fluctuations at the inner boundary lie in the range $1.08-1.61 \ {\rm km \ s^{-1}}$, which is consistent with observational constraints \citep{Oba_2020_ApJ}. We note that approximately 400 minutes of simulation time are required for $F^{\rm AW}_{\rm in}$ to converge to the prescribed value.

\subsection{Grid Setup and Numerical Solver}
In numerical models connecting the photosphere and solar wind, the use of a non-uniform grid is essential. This is due to the large scale gap between the lower atmosphere and the solar wind. In this study, we employ a grid system defined iteratively as follows.
\begin{align}
    \Delta r_{i+1} &= \max \Big[ \Delta r_{\rm min}, \ \Gamma (r_i; r_{\rm ge}) \Big], \\
    r_{i+1} &= r_i + \frac{1}{2} \left( \Delta r_i + \Delta r_{i+1} \right),
\end{align}
where $\Delta r_i$ and $r_i$ represent the size and radial distance of the center of the $i$-th grid, respectively, and
\begin{equation}
    \Gamma (r; r_{\rm ge}) = \min \left[ \Delta r_{\rm max}, \ \frac{2 \varepsilon_{\rm ge}}{2+ \varepsilon_{\rm ge}} \left( r -r_{\rm ge} \right) + \Delta r_{\rm min} \right]
\end{equation}
$\Delta r_{\rm min}$ and $\Delta r_{\rm max}$ represent the minimum and maximum grid sizes, set to $\Delta r_{\rm min} = 10 {\rm \ km}$ and $\Delta r_{\rm max} = 1000 {\rm \ km}$, respectively. $\varepsilon_{\rm ge}$, the rate of grid size expansion, is set to 0.01, while the grid expansion height, $r_{\rm ge}$, is fixed at $r_{\rm ge} = 1.04R_\odot$.

We have confirmed that Alfv\'en waves are well resolved with the above nonuniform grid system in almost all the cases. However, when the magnetic field becomes extremely weak in the corona, this model fails to resolve the Alfv\'en waves, leading to unphysical decline of the solar wind speed. In such cases, we adopt the following two-step expansion grid system:
\begin{align}
    &\Delta r_{i+1} = \max \Bigg[ \max \bigg[ \Delta r_{\rm min}, \ \Gamma_1 (r_i, r_{\rm ge,1}) \bigg], \ \Gamma_2 (r_i, r_{\rm ge,2}) \Bigg], \\
    &\Gamma_1 (r_i, r_{\rm ge,1}) = \min \Big[ \Delta r_{\rm mid}, \ \frac{2 \varepsilon_{\rm ge}}{2+ \varepsilon_{\rm ge}} \left( r_i -r_{\rm ge, 1} \right) + \Delta r_{\rm min} \Big], \nonumber \\
    &\Gamma_2 (r_i, r_{\rm ge,1}) = \min \left[ \Delta r_{\rm max}, \ \frac{2 \varepsilon_{\rm ge}}{2+ \varepsilon_{\rm ge}} \left( r_i -r_{\rm ge, 2} \right) + \Delta r_{\rm mid} \right], \nonumber 
\end{align}
where $\Delta r_{\rm min}=10 $ km, $\Delta r_{\rm mid}= 25-100$ km, $\Delta r_{\rm max}=1000$ km, $r_{\rm ge, 1}=1.04R_\odot$, $r_{\rm ge, 2}=2R_\odot$, $\varepsilon_{\rm ge}=0.01$.

In numerically solving Equations ~\eqref{eq:mass_conservation}--\eqref{eq:energy_conservation}, we adopted the cross-section-weighted variables, enabling the use of the Riemann solver as in the Cartesian coordinates \citep[see Section 2.8 of][]{Shoda_2021_AandA}. The finite volume method was employed to discretize the variables. The HLLD approximate Riemann solver \citep{Miyoshi_2005_JCoPh} was utilized for the calculation of the numerical flux at the cell boundary. In reconstructing the variables at cell boundaries, we used the 5th-order monotonicity-preserving scheme \citep{Suresh_1997_JCoPh, Matsumoto_2019_PASJ} for uniform grid regions and the 2nd-order MUSCL scheme \citep{van_Leer_1979_JCoPh} for non-uniform grid regions.

\section{Result}



In this study, we varied seven parameters listed in Table~\ref{tab:free parameters} ($B_{r,\odot}$ for Appendix) that regulate the flux-tube geometry to investigate how the solar wind properties depend on these parameters. Due to the large number of free parameters, it is difficult to distribute computational resources equally across all of them. Therefore, we first determined a set of typical parameters (listed in Table \ref{tab:free parameters}) and then focused on examining how the solar wind profile changes when some of these parameters are varied. These fiducial values and value ranges are determined with reference to the magnetic field extrapolation results obtained from the PFSS model. A list of the parameter combinations used in the analysis is provided in Table \ref{tab:simu_list_1} in Appendix. As mentioned above, these result can have the uncertainty of the velocity on the order of several tens of km s$^{-1}$.

\subsection{Dependence on $B_{r,\rm{cb}}$ and $f_{\infty}$} \label{sec:WS}
First, we discuss the effects of varying the coronal base magnetic field $B_{r,\rm{cb}}:=B_{r,\odot}/\eta_{\rm exp}$ and the asymptotic coronal expansion factor $f_{\infty}$ on the solar wind speed. Here, rather than organizing the description in terms of the direct control parameter $\eta_{\rm{exp}}$, we recast it in terms of $B_{r,\rm{cb}}$, which is more readily accessible as an observable coronal magnetic-field quantity. Specifically, we compare our simulation results with the WS empirical model, where the solar wind speed $v_{\rm WS}$ and $f_{\infty}$ are connected by the following relation \citep{Arge_2000_JGR}: 
\begin{align}
v_{\rm{WS}}= 267.5+\frac{410.0}{ {f_{\infty}}^{0.4}} \: \rm{km} \: \rm{s}^{-1}. \label{eq:WS}
\end{align}

\begin{figure}
  \begin{center}
  \plotone{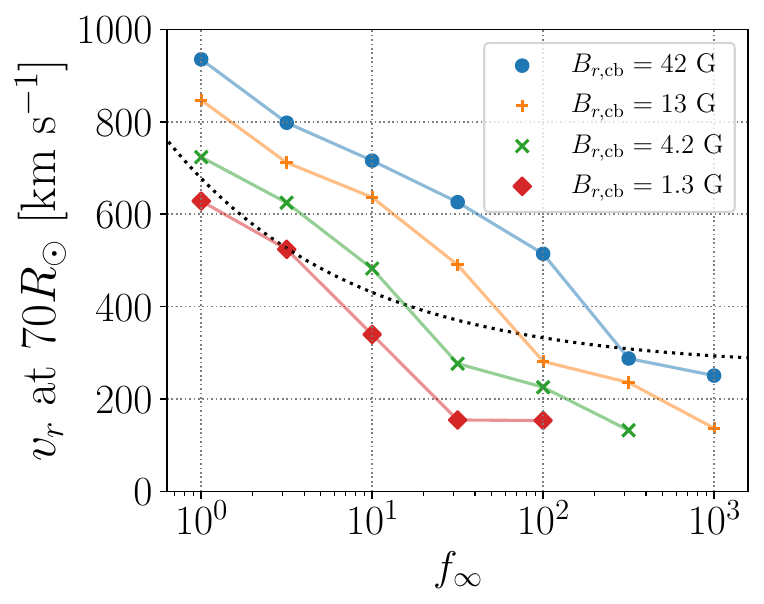}
  \caption{Relation between the asymptotic expansion factor $f_\infty$ and solar wind velocity at $r=70 R_{\odot}$. Simulation results correspond to the colored symbols and the WS model correspond to black dotted line. The results are limited for the case with only $f_\infty$ and $B_{r,\rm{cb}}$ parameters changed from default values. Each color and marker in the plot corresponds to a single value of $B_{r,\rm{cb}}$.
  \label{fig:WS}}
  \end{center}
\end{figure}
A comparison between our simulation results and the WS model is shown in Figure \ref{fig:WS}, where the colored symbols and black dotted line represent our simulation results and the WS model (Equation \eqref{eq:WS}), respectively. For simplicity, only the results obtained by varying $B_{r,\rm{cb}}$ and $f_{\infty}$ are shown here, while the results of surveys including variations of other parameters are presented in the upper-left panel of Figure \ref{fig:feature value}. When $B_{r,\rm{cb}}$ is fixed, a negative correlation between $f_{\infty}$ and the solar wind speed is observed. This result indicates that the present simulations do not significantly deviate from WS model. However, it should be noted that when $B_{r,\rm{cb}}/f_{\infty} \ge 0.42$ G ($B_{r,{\rm E}} > 0.91$ nT, left upper-left part of Figure \ref{fig:WS}), the results show an approximately linear trend for a fixed value of $ B_{r,\rm{cb}}$. In contrast, when $B_{r,\rm{cb}}/f_{\infty} \le 0.13$ G ($B_{r,{\rm E}} > 0.28$ nT or $v_r < 400$ km s$^{-1}$), the velocities are lower, and even with increasing $f_{\infty}$, there is a tendency for the velocity to no longer decrease further.

\subsection{Dependence on Expansion Height and Scale} \label{sec:exph}

\begin{figure*}[!t]
  \begin{center}
  \plotone{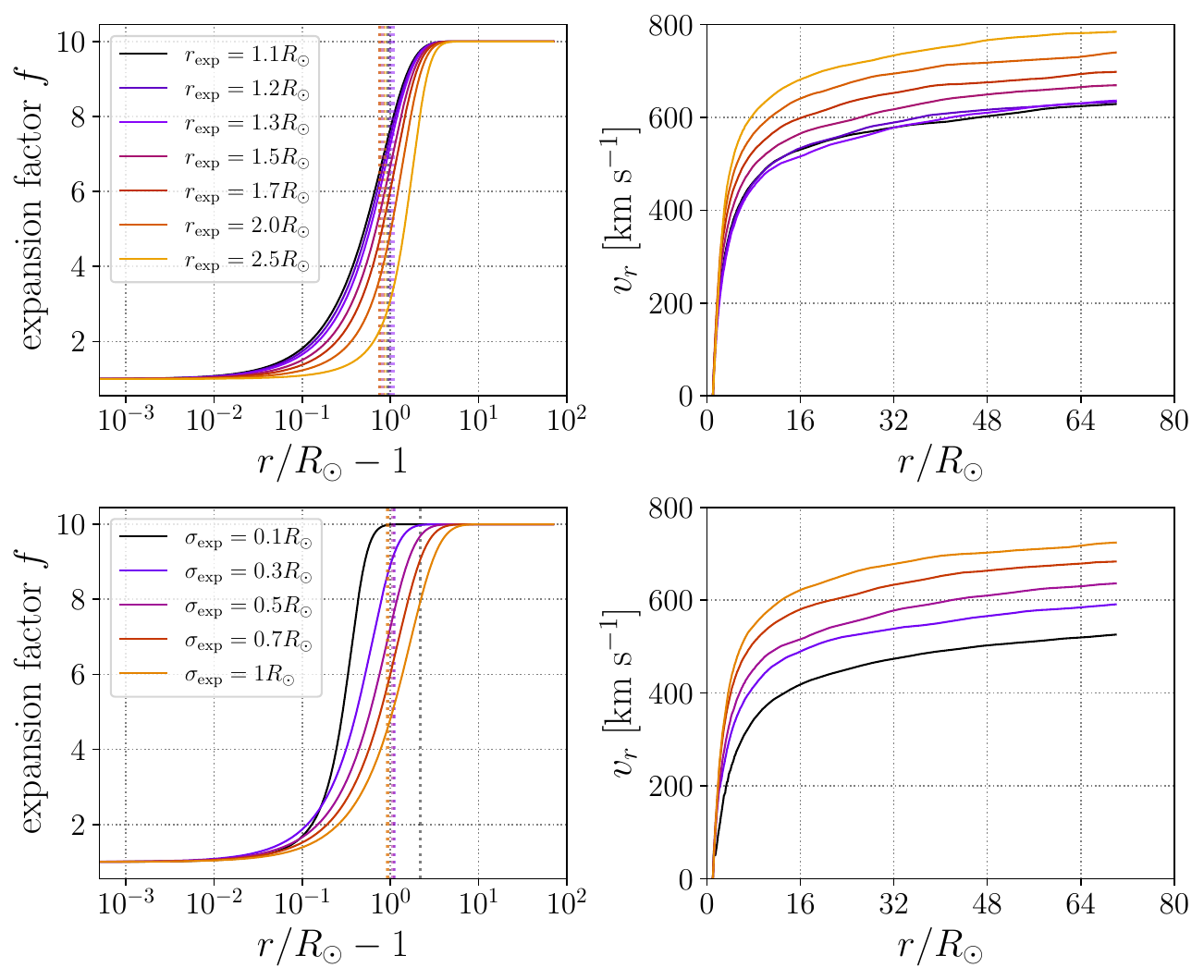}
  \caption{The line colors correspond to the same cases in the left and right panels. The upper and lower panels differ in the value of $r_{\rm{exp}}$ and $\sigma_{\rm{exp}}$, respectively: in the upper panels, $\sigma_{\rm{exp}}$ is fixed at $0.5R_\odot$, while in the lower panels, $r_{\rm{exp}}$ is fixed at $1.3R_\odot$. Left panels show the prescribed expansion factor profiles as a function of height from the photosphere $r/R_\odot-1$ (logarithmic scale). For reference, the transonic point in each simulation is indicated by a vertical dotted line. Right panels show the resulting solar wind speed profiles as a function of heliocentric distance $r$ (linear scale). In all cases presented here, the radial magnetic field strength at the coronal base, $B_{r,\rm{cb}}$, is fixed at $13$ G.
  \label{fig:exph_param}}
  \end{center}
\end{figure*}

Next, we present the simulation results of various the expansion height ($r_{\rm{exp}}$) and scale ($\sigma_{\rm{exp}}$). The upper panels of Figure \ref{fig:exph_param} show that fixing $\sigma_{\rm{exp}}$ at $0.5R_\odot$ and decreasing $r_{\rm{exp}}$ from $2.5R_\odot$ to $1.1R_\odot$ results in a $155$ km s$^{-1}$ reduction in the solar wind speed. This anti-correlation between $v_r$ and $r_{\rm{exp}}$ is qualitatively consistent with a previous observational study \citep{Dakeyo_2024_aap}. We note that, when $r_{\rm{exp}} \le 1.3R_\odot$, $f(r)$ and $v_r(r)$ show little differences. This is because, for sufficiently small $r_{\rm{exp}}$, the expansion factor $f(r)$ depends only on $\sigma_{\rm{exp}}$, rendering the solar wind speed insensitive to $r_{\rm{exp}}$. From this result, it becomes clear that it is not the value of $r_{\rm{exp}}$ determined by the condition $f'' = 0$—as discussed in \citet{Dakeyo_2024_aap}—but rather the overall profile of $f(r)$ that influences the solar wind speed. The lower panels of Figure \ref{fig:exph_param} shows that, when $r_{\rm{exp}}$ is fixed at $1.3R_\odot$ and $\sigma_{\rm{exp}}$ is varied from $1R_\odot$ to $0.1R_\odot$, the solar wind speed decreases by $198$ km s$^{-1}$. As seen in Figure \ref{fig:WS}, producing a comparable difference in velocity requires varying $f_\infty$ by approximately an order of magnitude.

\subsection{Effect of Non-monotonic Expansion}

According to the results of magnetic field extrapolation, some open flux tubes, in particular those near pseudostreamers, are likely to undergo the non-monotonic expansion \citep{Panasenco_2013_AIPC}. Therefore, alongside the expansion factor and the height or scale of expansion, non-monotonicity is a key parameter characterizing open flux tubes. Bearing this fact in mind, we present the results for cases in which magnetic flux tubes in the corona exhibit non-monotonic expansion. Specifically, we varied $g_{\rm{max}}, r_{\rm{max}},$ and $w$ to investigate the dependence of the wind speed on these parameters. First, we assess the effect of non-monotonic expansion by increasing $g_{\rm{max}}$, while keeping $r_{\rm{max}}$ and $w$ constant. Then, we examine the influence of varying $r_{\rm{max}}$ and $w$.

\begin{figure*}[!t]
  \begin{center}
  \plotone{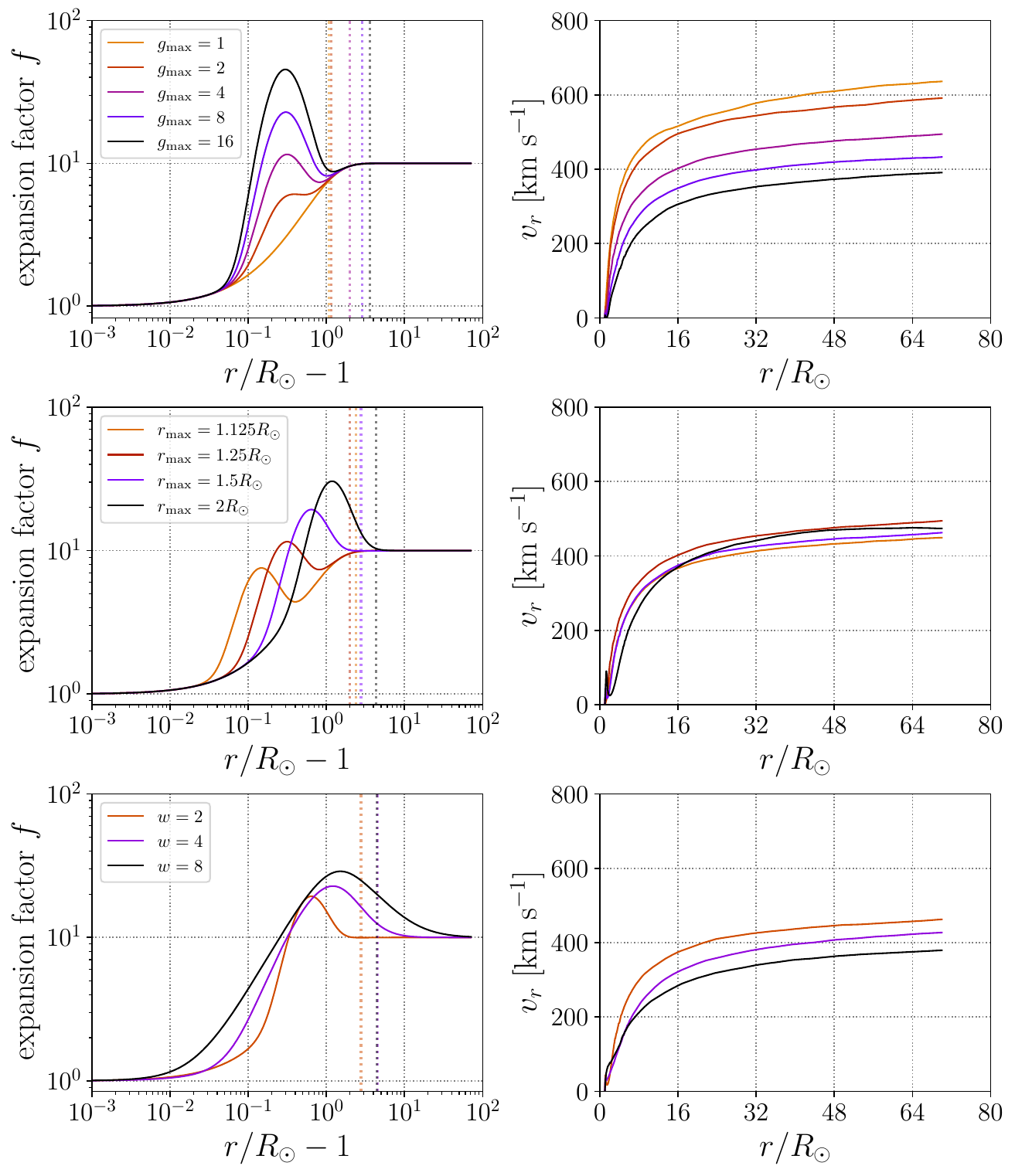}
  \caption{The same as Figure \ref{fig:exph_param} but the cases differ in $g_{\rm{max}}$(top panels), $r_{\rm{max}}$(middle panels), $w$(bottom panels)
  \label{fig:nonmono_param}}
  \end{center}
\end{figure*}

As shown in the top panels of Figure \ref{fig:nonmono_param}, increasing $g_{\rm{max}}$ from 1 leads to a decrease in the solar wind speed. Specifically, $v_r$ measured at $r = 70R_\odot$ decreased by up to $245$ km s$^{-1}$, suggesting that non-monotonicity is also an important factor on solar wind speed. In the middle panel, we compare the cases where $g_{\rm{max}}=4$ and $w=2$ are fixed, and $r_{\rm{max}}$ varied as $1.125R_\odot,\ 1.25R_\odot,\ 1.5R_\odot,\ 2R_\odot$.. The maximum difference in $v_r$ among these cases was approximately $45.1$ km s$^{-1}$, which means that the effect of $r_{\rm{max}}$ is relatively small. However, the variations are not significant enough to exhibit a systematic trend with $r_{\rm{max}}$. While this result may appear to contradict the previous finding in Section \ref{sec:exph} that the expansion height has a significant impact on the solar wind speed, it cannot be regarded as necessarily contradictory considering that changing $r_{\rm{max}}$ affects not only the location of the "hump" in $f(r)$ but also the absolute value of $f$ at the hump. A smaller $r_{\rm{max}}$ lowers the radial height of the hump, which tends to reduce the solar wind speed. On the other hand, since $g_{\rm{max}}$ is fixed, the value of $f$ at the peak of the hump decreases, which increases the wind speed. 

The bottom panels compare the cases where $g_{\rm{max}}$ and $r_{\rm{max}}/R_\odot$ are fixed to 4 and 1.5, respectively, and $w$ is varied among 2, 4, and 8. In the left panel, changes in the height and location of the hump in $f(r)$ are observed. This is due to the fact that ${\rm d}f(r)/{\rm d} r$ depends not only on the Gaussian profile but also on the basal component $f_{\rm{KH}}(r)$ (which increases monotonically with $r$), as given by Equation \eqref{eq:f}. The variation in $w$ results in a maximum difference in the solar wind speed of $83$ km s$^{-1}$. When $w=4$ and $w=8$, $f(r)$ converges to its asymptotic value at relatively high altitudes. Given that the source surface in magnetic field extrapolations is typically placed between $1.5-3.5$ $R_\odot$ \citep{Hoeksema_1983_JGR, Arden_2014_JGRA, Benavitz_2024_ApJ}, these settings may be considered unrealistic. As shown in Section \ref{sec:exph}, the value of $f(r)$ at lower altitudes has a strong influence on the solar wind speed. Thus, the velocity differences caused by the variations in $w$ are likely due to the increase in $f(r)$ in the low-altitude region ($r \sim 1.01R_\odot$–$1.3R_\odot$).

\subsection{New Features for Solar Wind Speed}

From the simulation results presented so far, we infer that the detailed radial profile of the flux-tube cross section plays a crucial role in determining the solar wind speed. As discussed in Section \ref{sec:exph}, it is the resulting functional form of the expansion factor $f(r)$, rather than parameters such as $r_{\rm{exp}}$ and $\sigma_{\rm{exp}}$ that characterize the expansion profile, that plays a critical role in determining the solar wind speed in this model. Therefore, physical quantities characterizing the solar wind speed are likely to be better represented by the radial profile of the expansion factor $f(r)$ from the coronal base to a certain height, rather than by a single-point value such as $f_\infty$.

\begin{figure*}[!t]
  \begin{center}
  \plotone{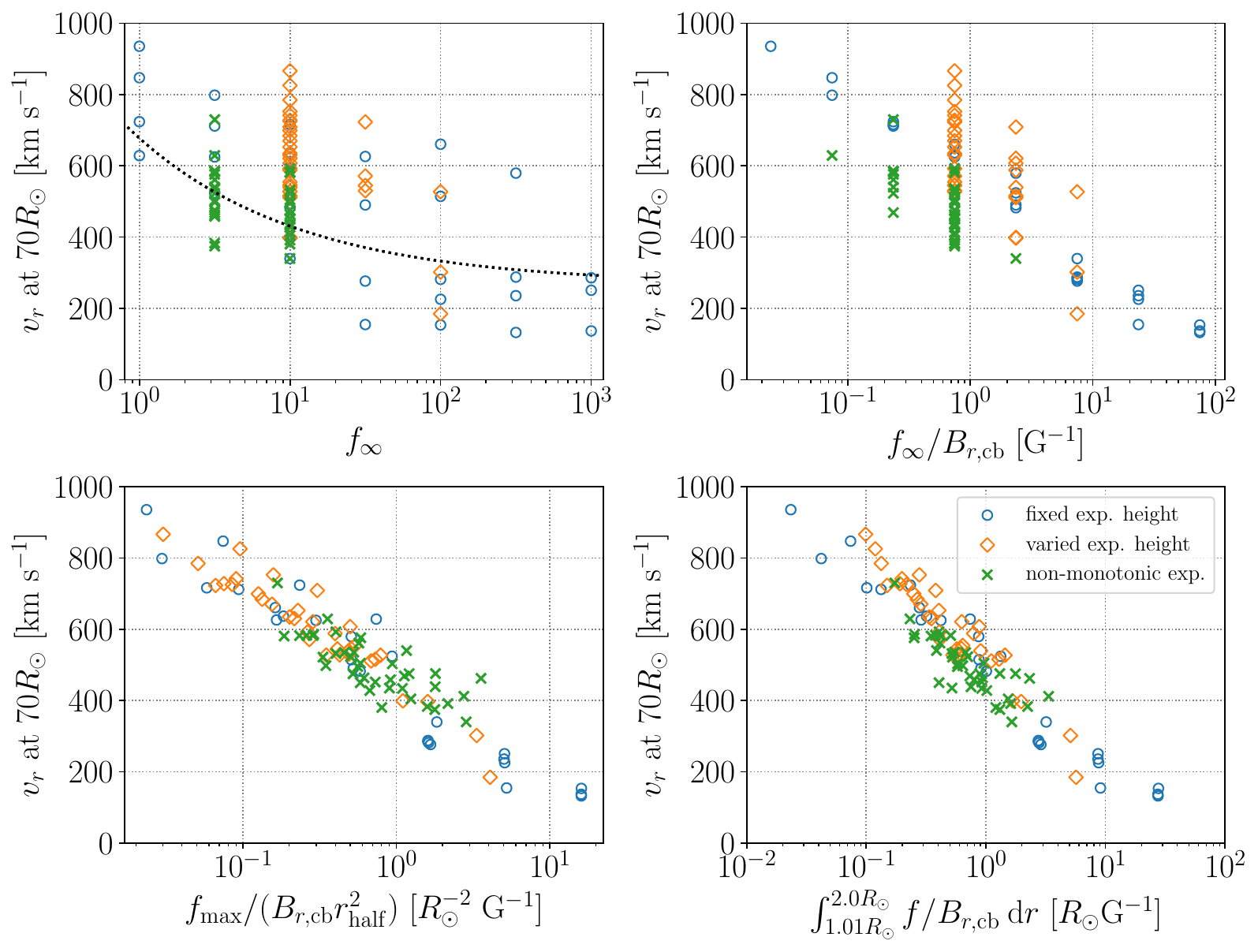}
  \caption{Relation between wind velocity measured at $r=70 R_\odot$ and each feature candidate. Upper left: asymptotic expansion factor $f_{\infty}$. The wind speed predicted by the Wang-Sheeley model $v_{\rm{WS}}$ is indicated by the dotted line for comparison. Upper right: the quantity $f_{\infty}/B_{r,\rm{cb}}$, where $B_{r,\rm{cb}}$ is the radial magnetic field strength at the coronal base. Lower left: the quantity $f_{\rm{max}}/(B_{r,\rm{cb}}r_{\rm{half}}^2)$, where $f_{\rm{max}}$ is the maximum value of $f(r)$ and $r_{\rm{half}}$ is the radial distance at which $f(r)=f_{\rm{max}}/2$ for the first time. Lower right: integral value of $f(r)/B_{r,\rm{cb}}$ from coronal base to a certain height in atmosphere. Cases are divided into three groups: (1) monotonic expansion for fixed expansion height and scale($r_{\rm{exp}}$ \& $\sigma_{\rm{exp}}$)(blue circles) (2) monotonic expansion for varied expansion height and scale(orange diamonds) (3) non-monotonic expansion, whose $g_{\rm{max}}$ is larger than $1$(green crosses).
  \label{fig:feature value}}
  \end{center}
\end{figure*}

\begin{deluxetable}{lccc}
\tablecaption{List of the Pearson correlation coefficient for each category corresponding to each feature value. The labels of the category are the same as Figure \ref{fig:feature value}. \label{tab:corrcoef}}
\tablewidth{0pt}
\tablehead{
  \colhead{} & \multicolumn{3}{c}{correlation coefficient with} \\[2pt]
  \colhead{feature} & \colhead{(1)} & \colhead{(2)} & \colhead{(3)}
}
\startdata
$f_\infty$                                        & $-0.76$ & $-0.54$ & $-0.32$ \\
$f_\infty/B_{r,\rm{cb}}$                          & $-0.98$ & $-0.67$ & $-0.62$ \\
$f_{\rm{max}}/(B_{r,\rm{cb}}r_{\rm{half}}^2)$     & $-0.97$ & $-0.93$ & $-0.82$ \\
$\displaystyle \int_{1.01R_\odot}^{2.0R_\odot}f/B_{r,\rm{cb}}\,\dd{r}$ & $-0.97$ & $-0.95$ & $-0.83$ \\
\enddata
\end{deluxetable}

Based on the preceding discussion, we compare the predictive performance of candidate feature values that incorporate the radial profile of the expansion factor $f(r)$ with those that do not. Figure \ref{fig:feature value} consists of four panels showing the relation between solar wind velocity and each feature value. For visualization, all of the cases were grouped into three categories: (1) monotonic expansion cases with fixed expansion height and scale, (2) monotonic expansion cases with various expansion heights and scales, and (3) non-monotonic expansion cases. For a fair comparison, we calculate the Pearson correlation coefficient (PCC) between $\left.v_r\right|_{\rm out}$ and the logarithmic of each feature value for each category, and summarize the results in Table \ref{tab:corrcoef}. 

The upper left panel of Figure \ref{fig:feature value} extends the data presented in Figure \ref{fig:WS} by incorporating the results on expansion height and scale and non-monotonicity, yet it still exhibits a negative correlation  between $v_r$ and $f_\infty$. This trend is consistent with the observational results \citep{Wang_1990_ApJ, Wang_2020_ApJ}. Table \ref{tab:corrcoef} shows that $f_\infty$ exhibits the weakest correlation among the four feature values across all categories. 

The upper right panel of Figure \ref{fig:feature value} presents the relationship between $v_r$ and $f_\infty / B_{r,\rm{cb}}$—the ratio of the asymptotic expansion factor to the coronal base magnetic field. We note that the quantity $f_{\infty}/B_{r,\rm{cb}}$ corresponds to the source-surface magnetic field strength in the PFSS extrapolation \citep{Altschuler_1969_SoPh, Schatten_1969_SoPh}. Specifically, for the source-surface height of $2.5R_\odot$, magnetic flux conservation gives $f_{\infty}/B_{r,\rm{cb}} = 6.25/B_{r,\rm{SS}}$, where $B_{r,\rm{SS}}$ denotes the radial field at the source surface. When considering only (1) monotonic expansion for fixed expansion height and scale cases, a strong negative correlation with wind velocity is observed (PCC: $-0.98$). However, correlations get weak with (2)variation in expansion height and scale (PCC: $-0.67$) and (3) non-monotonic expansion cases  (PCC: $-0.62$). 

Meanwhile, the features which can reflect the middle-altitude region of $f(r)$ has strong correlations with all of the groups as lower panels of Figure \ref{fig:feature value}. The feature value of the lower left panel is the quantity $f_{\rm{max}}/(B_{r,\rm{cb}}r_{\rm{half}}^2)$, where $f_{\rm{max}}$ is the maximum value of $f(r)$ and $r_{\rm{half}}$ is the minimum radial distance that satisfies $f(r)=0.5 f_{\rm{max}}$. We introduce $r_{\rm{half}}$ as a proxy for the expansion height and scale, since actual open flux tubes are generally not well represented by Equations \eqref{eq:A}–\eqref{eq:rg}, and a more robust parameter is needed to apply our results to heliospheric modelling. 

The feature value of the lower right panel is the integral of $f(r)/B_{r,\rm{cb}}$ with respect to $r$ from $1.01R_\odot$ to $2.0R_\odot$, $\int_{1.01R_\odot}^{2.0R_\odot}f(r)/B_{r,\rm{cb}} \dd{r}$. Notably, the feature values in the lower panels of Figure \ref{fig:feature value} exhibit similar trends; they show relatively strong correlations across all categories, and particularly high correlations in the monotonic expansion cases (categories (1) and (2)). The lower limit of the integration range is set slightly above the coronal base. The upper limit is tuned so that the integrated quantity has the largest correlation with the wind speed. Interestingly, it is found to be close to the conventional source-surface height or the sonic point. We note, however, that the correlation is not highly sensitive to the choice of the integration range (see Appendix \ref{app:integration region}).

\section{Discussion}

\subsection{Physical Origin of the Wind Speed Variation}

We verify the validity of the model used in this study. The numerical model employed here has been validated in \citet{Shoda_2022_ApJ}, and, as shown in Section \ref{sec:WS}, it does not contradict the observational trends described by the WS model. However, since this study artificially introduces special forms of the expansion factor $f(r)$ such as non-monotonic expansion, it is necessary to confirm that the present results are not inconsistent with observed pseudostreamer velocities. In the present simulation, the solar-wind speed at $r = 70R_\odot$ for the non-monotonic expansion cases ranges between $300-600$ km s$^{-1}$. It should be noted, however, that although these cases are constructed with reference to the pseudostreamer $f$ values derived from the PFSS model, some of them represent rather extreme configurations. Previous in situ measurements at 1 au, such as those by \citet{Riley_2012_SoPh} and \citet{Wang_2012_ApJ} using ACE data, reported pseudostreamer speeds of approximately $300–550$ km s$^{-1}$, while \citet{Wang_2019_ApJ} reported $300–600$ km s$^{-1}$ from OMNI observations at 1 au. Closer-in measurements, such as those by \citet{D'Amicis_2021_A&A} with Solar Orbiter at $0.65$ au, indicate pseudostreamer speeds of about $300–450$ km s$^{-1}$. Furthermore, \citet{Badman_2023_JGR} reported PSP measurements near $r = 13R_\odot$ showing speeds of $150–300$ km s$^{-1}$; similar speeds are reproduced in our simulation at $r = 13R_\odot$ for cases with large $g_{\rm{max}}$ (Figure \ref{fig:nonmono_param}). From these comparisons, we conclude that the present model does not exhibit any significant deviation from the observed pseudostreamer wind speeds.

To additionally confirm that the present results are physically grounded, we focus on the following elements of the expansion factor and confirm through numerical simulations that each of them influences the solar wind speed: (a) the magnitude of the asymptotic expansion factor $f_\infty$, (b) the magnetic field strength at the coronal base ($B_{r,\rm{cb}}=B_{r,\odot}/\eta_{\rm{exp}}$), (c) the expansion height and scale ($r_{\rm{exp}},\sigma_{\rm{exp}}$) and (d) the non-monotonic expansion ($g_{\rm{max}}, \ r_{\rm{max}},\ w$). In the following, we discuss the physical mechanisms by which each parameter affects the solar wind acceleration.

The radial dependence of the steady solar wind speed $v_r(r)$ can be described as follows by the radial momentum balance under the assumptions of an approximately isothermal plasma and the dominance of Alfvén waves over other wave modes.
\begin{align}
&\qty(1-\frac{C_s^2}{v_r^2})v_r \pdv{v_r}{r} \notag  \\ 
&=C_s^2\qty(\frac{2}{r}+\dv{r} \ln f)-\frac{1}{8\pi \rho}\pdv{B_\perp^2}{r}-\frac{GM_\odot}{r^2}, \label{eq:radial momentum}
\end{align}
where $C_s$ denotes the sound speed (i.e., proportional to the square root of temperature $T$). In all simulations, the variation in $C_s$ within the subsonic region stays within several tens of km s$^{-1}$. Accordingly, the solar wind speed at the sonic point (equal to the sound speed) is nearly identical across all cases, indicating that the difference in the asymptotic wind speed arises from processes in the supersonic region.

In the supersonic region, the term in parentheses on the left-hand side of Equation~\eqref{eq:radial momentum} is positive; therefore, a larger right-hand side leads to a higher solar wind speed. Since the gravitational term is identical across all cases, the varying parameters are $C_s$, ${\rm d} \ln f/ {\rm d}r$, and $\partial B_\perp^2/ \partial r$. Hereafter, we simplify the analysis by assuming $\partial B_\perp^2 / \partial r \propto B_\perp^2$ in the supersonic region, corresponding to Alfvén wave decay with a fixed length scale. As flux-tube expansion is typically negligible in the supersonic region, except for cases with non-monotonic or high-altitude (cf. Figures~\ref{fig:exph_param}, \ref{fig:nonmono_param}), long-scale expansion, the contribution from ${\rm d} \ln f/ {\rm d}r$ can usually be ignored. The asymptotic wind speed is then primarily controlled by $C_s$ and $B_\perp^2/\rho$ integrated over the supersonic region.

\begin{figure}[!t]
  \begin{center}
  \plotone{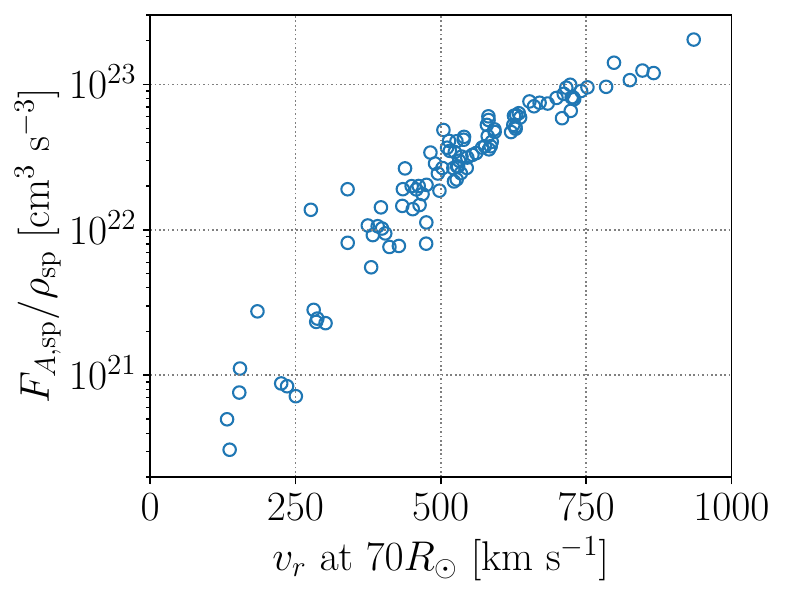}
  \caption{Relation between wind velocity $v_r$ at $r=70R_\odot$ and Alfvén wave energy flux per mass $F_A/\rho$ at the sonic point. 
  \label{fig:FA_vr}}
  \end{center}
\end{figure}

We hypothesize that $C_s$ and $B_\perp^2/\rho$ in the supersonic region strongly correlate with the Poynting flux per unit mass measured at the sonic point ($F_{A,{\rm sp}}/\rho_{\rm sp}$), where $F_A$ is given by Equation~\eqref{eq:def F_A} and $X_{\rm sp}$ denotes that value of $X$ at the sonic point. The reason for this hypothesis is as follows. Since the heat source is Alfv\'en wave energy, the temperature (or sound speed) in the acceleration region is determined by the injected energy flux per unit mass density ($F_A/\rho$). The normalization by mass is essential, as a greater number of particles requires more energy to raise the temperature. A similar argument applies to $B_\perp^2/\rho$, since in the Alfv\'enic regime, the energy flux is approximated by
\begin{equation}
    \frac{B_\perp^2}{\rho} \approx \frac{F_A}{\rho \left( V_A + 3v_r/2 \right)}, \label{eq:energy_flux_per_unit_mass_density}
\end{equation}
where we use $\boldsymbol{v}_\perp \approx - \boldsymbol{B}_\perp / \sqrt{4 \pi \rho}$ and $V_A \approx B_r / \sqrt{4 \pi \rho}$.
Equation~\eqref{eq:energy_flux_per_unit_mass_density} thus indicates a positive correlation between $B_\perp^2/\rho$ and $F_A/\rho$, implying that a larger $F_A/\rho$ at the sonic point leads to a higher $B_\perp^2/\rho$ in the supersonic region. Consequently, $F_A/\rho$ at the sonic point should correlate positively with the asymptotic wind speed, as shown in Figure~\ref{fig:FA_vr}. We therefore conclude that the wind speed variation is governed by $F_A/\rho$ at the sonic point.

\begin{figure*}[!t]
  \begin{center}
  \plotone{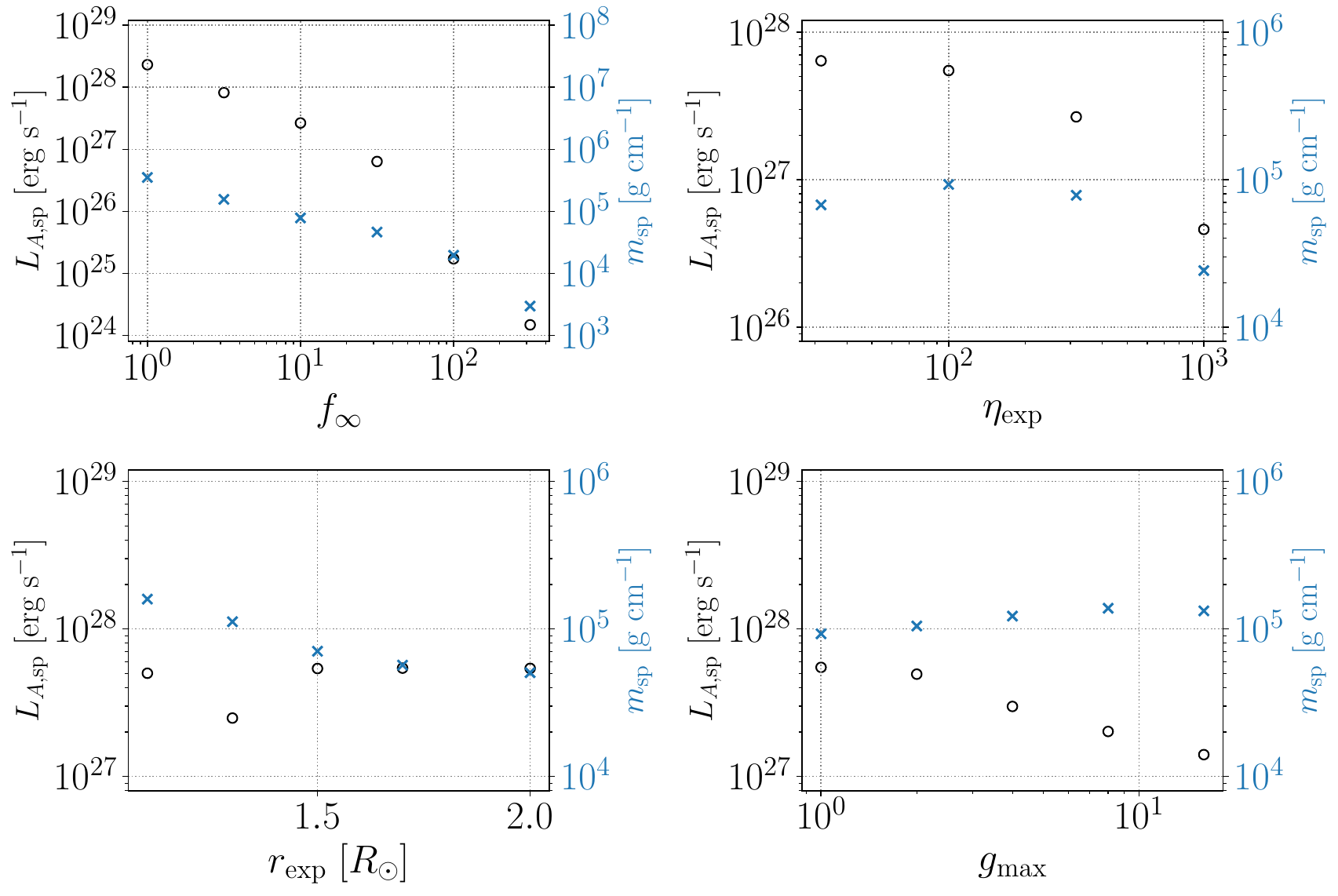}
  \caption{Alfvén wave energy flux $L_A \propto F_A A$ and mass per cross section $m \propto \rho A$ at the sonic point as functions of each parameter. Upper left: dependence on (a) $f_\infty$, with $B_{r,\mathrm{cb}} = 42$ G fixed. Upper right: (b) dependence on $\eta_{\rm exp}$, with $f_\infty = 10$ G fixed. Lower left: (c) dependence on $r_{\mathrm{max}}$, with $\sigma_{\mathrm{max}} = 0.1R_\odot$ fixed. Lower right: (d) dependence on $g_{\mathrm{max}}$, with $r_{\mathrm{max}} = 1.25R_\odot$ and $w = 2$ fixed.
  \label{fig:FAandrho}}
  \end{center}
\end{figure*}

In Figure \ref{fig:FAandrho}, we show the dependence of $F_{A,{\rm sp}}$ and $\rho_{\rm sp}$ on the flux-tube parameters.
Here, the plotted quantities $L_A$ and $m$ are defined as follows: 
\begin{align}
L_A&=4\pi r^2 \frac{\eta f}{\eta_{\rm{exp}} f_\infty}F_A \, ,\\
m&=4\pi r^2 \frac{\eta f}{\eta_{\rm{exp}} f_\infty}\rho \, .
\end{align}
We note that $\eta f/\eta_{\rm{exp}} f_\infty$ represents the filling factor and $F_A/\rho \equiv L_A/m$, and, instead of the original $F_A$ and $\rho$, we show the quantities weighted by the flux tube cross section. $L_{A,{\rm{sp}}}$ and $m_{\rm{sp}}$ are used in place of $F_{A,{\rm{sp}}}$ and $\rho_{\rm{sp}}$ as $L_{A,{\rm{sp}}}$ and $m_{\rm{sp}}$ do not decrease radially with flux-tube expansion and are less influenced by changes in the sonic-point radius.

As seen in the upper left panel of Figure \ref{fig:FAandrho}, increasing $f_\infty$ causes a rapid decrease in $L_A$. In the regime $f_\infty \le 10^{1.5}$, one finds $L_{A,{\rm{sp}}} \propto f_\infty^{-1}$, reflecting that the incident energy flux from the photosphere remains fixed while the effective cross‐sectional area of the flux tube at the photosphere shrinks as $f_\infty$ grows. In other words, the decrease with $f_\infty$ reflects the reduction in the Alfvén speed, $V_A = B_r/\sqrt{4\pi \rho}$, whereas $v_\perp$ and $B_\perp$ remain nearly unchanged. For $f_\infty \ge 10^2$, $L_{A,{\rm{sp}}}$ falls even more steeply. In contrast, $m_{\rm{sp}}$ exhibits a much smaller variation than $L_{A,{\rm{sp}}}$, which can be attributed to the subsonic (gravitationally stratified) region where the density $\rho$ is mainly by the coronal temperature. 

As shown in the upper right panel of Figure \ref{fig:FAandrho}, $m_{\rm{sp}}$ shows a smaller and non-monotonic variation with $\eta_{\rm{exp}}$, in contrast to $L_{A,{\rm{sp}}}$ which varies more significantly. This trend is similar to the case of $f_\infty$. An decrease in $\eta_{\rm exp}$, that is, a increase in $B_{r,\rm cb}$, has two competing effects on $F_{A,\rm{cb}}$; it tends to enhance $F_{A,\rm{cb}}$ by decreasing the chromospheric expansion, while it also tends to reduce $F_{A,\rm{cb}}$ because a larger Alfv\'en speed gap at the transition region enhances wave reflection. As a result, this study finds a scaling of $F_{A,\rm{cb}} \propto \eta_{\rm{exp}}^{-0.42}$, although the correlation is weak. In contrast, the correlation between $\eta_{\rm{exp}}(\propto 1/B_{r,\rm{cb}})$ and $v_r$ is significant, as shown in Figure \ref{fig:WS}. These findings indicate that the higher wind speed associated with smaller $\eta_{\rm exp}$ and larger $B_{r,\rm cb}$ is not due to increased energy flux at the coronal base, but rather to another physical mechanism.

A possible interpretation of the faster solar wind with $B_{r,\rm cb}$ is that turbulent heating in the subsonic region is suppressed when $B_{r,\rm cb}$ is large. Qualitatively, as $B_{r,\rm cb}$ increases, the Alfvén-wave propagation time becomes shorter relative to the eddy turnover time, leading to the suppression of turbulence. In what follows, we demonstrate that the phenomenological turbulence model employed in this simulation exhibits the same trend. Assuming $F_{A,\rm{cb}}$ is independent of $B_{r,\rm cb}$, the outward Els\"asser variable at the coronal base is approximated by
\begin{equation}
    z_{\perp,{\rm cb}}^+ \approx \left( \frac{F_{A,{\rm cb}}}{\rho_{\rm cb} V_{A,{\rm cb}}} \right)^{1/2} \propto \rho_{\rm cb}^{-1/4} B_{r,{\rm cb}}^{-1/2}.
\end{equation}
Given that the inward Els\"asser amplitude is proportional to the outward one, the turbulent heating rate at the coronal base becomes
\begin{align}
    Q_{\rm turb, cb} &= \frac{c_d}{4 \lambda_{\perp,{\rm cb}}} \rho_{\rm cb} \left( {z_{\perp,{\rm cb}}^+}^2 z_{\perp,{\rm cb}}^- + {z_{\perp,{\rm cb}}^-}^2 z_{\perp,{\rm cb}}^+ \right) \notag \\
    &\propto \rho_{\rm cb} \frac{{z_{\perp,{\rm cb}}^+}^3}{\lambda_{\perp,{\rm cb}}} \propto \rho_{\rm cb}^{-1/4} B_{r,{\rm cb}}^{-1}, \label{eq:turbulent_heating_scaling}
\end{align}
where we use $\lambda_{\perp,{\rm cb}} \propto B_{r,{\rm cb}}^{-1/2}$. Considering that the dependence on $\rho$ in Equation~\eqref{eq:turbulent_heating_scaling} is weaker, a larger $B_{r,\rm{cb}}$ results in a lower heating rate at the base. Although this trend is not guaranteed to hold throughout the entire subsonic region, it is reasonable to expect a similar $B_{r,\rm{cb}}$ dependence.

As shown in the lower left panel of Figure \ref{fig:FAandrho}, $L_{A,{\rm{sp}}}$ is almost unchanged by the expansion height, whereas $m_{\rm{sp}}$ declines as the expansion height increases. As noted by \citet{Dakeyo_2024_aap}, varying the expansion height (and scale) affects whether ${\rm d}f/{\rm d}r$ increases in the subsonic region, where the bracket on the left-hand side in Equation \eqref{eq:radial momentum} becomes negative, or in the supersonic region, where the bracket on the right-hand side becomes positive. In other words, when the expansion occurs at lower altitudes, the flux tube expansion works more to decelerate the solar wind, resulting in a lower wind speed. This likely explains the wind speed dependence on $r_{\rm exp}$ (and $\sigma_{\rm exp}$).

\begin{figure}[!t]
  \begin{center}
  \plotone{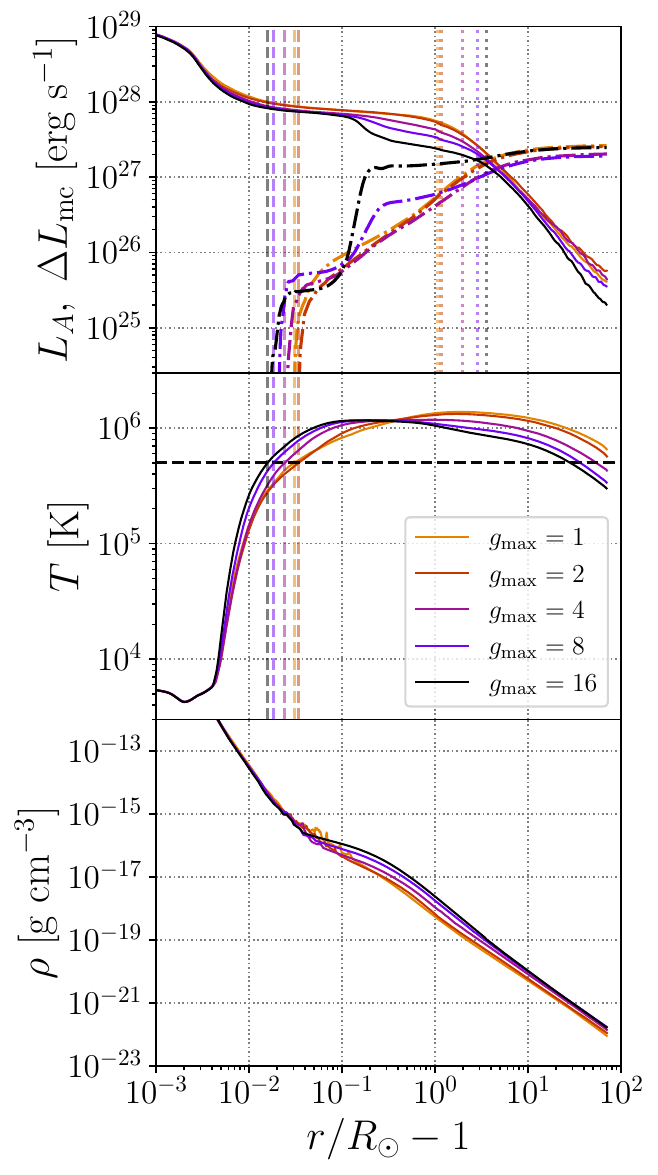}
  \caption{Profiles corresponding to the cases displayed in the top panel of Figure \ref{fig:nonmono_param}. Top panel: Alfvén-wave energy flux $L_A(r)$ (solid line) and the cumulative energy converted by the mode conversion from the coronal base $r_{\rm cb}$ up to each $r$, $\Delta L_{\rm mc}(r)$ (dot-dash line). The coronal base $r_{\rm cb}$ is defined as the height where $T$ first exceeds $5 \times 10^{5}$ K. The resulting $r_{\rm cb}$ positions are marked by vertical dashed lines in both the top and middle panels. For reference, the transonic point in each simulation is indicated by a vertical dotted line. Middle panel: temperature $T(r)$. The criteria for $r_{\rm cb}$, $5 \times 10^{5}$ K, is plotted as horizontal black dashed line.  Bottom panel: mass density $\rho (r)$.
  \label{fig:LA_nonmono}}
  \end{center}
\end{figure}

In the case of non-monotonic expansion (lower right panel of Figure \ref{fig:FAandrho}), in addition to the usual expansion effects, a local decrease in $B_r$ enhances wave dissipation and increases the temperature in the subsonic region. To investigate the origin of this enhanced dissipation, Figure \ref{fig:LA_nonmono} presents the radial profiles of the Alfv\'en wave energy flux $L_A(r)$, the wave energy loss rate by work $\Delta L_{\rm{mc}}(r)$, the transonic point, temperature $T(r)$, and density $\rho(r)$ for the same cases shown in the top panel of Figure \ref{fig:nonmono_param}. The quantity $\Delta L_{\rm{mc}}$ is defined as \citep{Shimizu_2022_ApJ}:
\begin{align}
    \Delta L_{\rm{mc}}&=\int_{r_{\rm{cb}}}^r 4\pi r^2 \frac{\eta f}{\eta_{\rm{exp}} f_\infty}Q_{\rm{mc}} \dd{r} \, , \\
    Q_{\rm mc}&=\qty(\rho v_\perp^2-\frac{B_\perp^2}{4\pi})\frac{v_r}{2}\dv{\ln A}{r}-v_r\pdv{r}\qty(\frac{B_\perp^2}{8\pi}) \, .
\end{align}
We note that the work done by Alfv\'en waves $Q_{\rm mc}$ includes both solar wind acceleration via wave pressure \citep{Belcher_1971_ApJ} and local generation of slow-mode waves through nonlinear mode conversion \citep{Kudoh_1999_ApJ, Suzuki_2005_ApJ}. We also note that the definition of $Q_{\rm mc}$ does not directly represent the generation of slow-mode waves. However, in one-dimensional solar-wind MHD models, it is well established that large-amplitude slow-mode waves are generated, and that these waves arise from local plasma compression driven by wave pressure \citep[including parametric decay,][]{Shoda_2018_ApJ_frequency_dependent}. Therefore, the effects of the generation of slow-mode waves are assessed through $Q_{\rm mc}$. In the top panel, for cases with extreme weakening of $B_r$ near $r = r_{\rm{max}} = 1.25R_\odot$, such as $g_{\rm{max}} = 8$ and $16$, $L_A$ decreases around $r = r_{\rm{max}}$. This decrease in $L_A$ in the subsonic region corresponds to the work done by Alfv\'en wave, denoted as $\Delta L_{\rm{mc}}$. This work is likely caused by mode conversion, triggered by enhanced wave nonlinearity due to flux tube expansion and weak $B_r$. The resulting increase in the density scale height in the subsonic corona raises the mass density at the sonic point \citep{Holzer_1980_JGR}, thereby reducing the solar wind speed.

\subsection{Limitation and Possible Applications}

In this study, we find that the way open flux tubes expand with radial distance significantly influences the solar wind speed. We note, however, that care must be taken when assessing consistency with actual observations. To improve the accuracy, our model should be calibrated against three-dimensional models that can resolve wave generation \citep{Finley_2022_AandA, Kuniyoshi_2023_ApJ} and turbulence development \citep{van_Ballegooijen_2016_ApJ, Chandran_2019_JPlPh, Shoda_2019_ApJ}. In addition, interchange reconnection and small-scale jets, suggested by earlier studies \citep{Wang_2020_ApJ, Bale_2023_Natur, Chitta_2023_Sci, Iijima_2023_ApJ} as possible contributors to solar wind acceleration, are not included in the present analysis. Moreover, it is important to evaluate whether the value of the energy flux incident into the corona, $F_{A,\rm{cb}}$, is realistic, since $F_{A,\rm{cb}}$ is a crucial factor for the solar wind properties. It is necessary to compare relationships in this simulation with actual solar observations and update the model accordingly. 

For simplicity, we considered a straight flux tube extending radially. In reality, however,open flux tubes are frequently non-radial, and their curvature plays a significant role in solar wind dynamics \citep{Li_2011_A&A, Pinto_2016_A&A}. Therefore, our findings should be carefully re-evaluated in light of curvature effects. For instance, while we find that $\int f/B_{r,\rm{cb}}\dd{r}$ showed the best correlation, it is likely that the correct integral should be over $\dd{s}$ rather than $\dd{r}$. In this regard, it is necessary to perform validation using more realistic open flux tube geometries based on extrapolated magnetic fields from actual magnetograms, as done by \citet{Shoda_2022_ApJ}.

Despite the aforementioned limitations, our results suggest that the insights gained from the physics-based model can contribute to improving the accuracy of space weather prediction. In particular, the characteristic quantities proposed in this study, such as $f_{\rm{max}}/(B_{r,\rm{cb}}r_{\rm{half}}^2)$ and $\int_{1.01R_\odot}^{2.0R_\odot}f/B_{r,\rm{cb}},\dd{r}$, are extensions of $f_{\infty}/B_{r,\rm{cb}}$ proposed by \citet{Fujiki_2005_AdSpR} and \citet{Suzuki_2006_ApJ}, with integration and specific parameters introduced to incorporate information such as non-monotonic expansion and variation in expansion height or scale. By taking these factors into account, it is suggested that the uncertainty of up to $\pm 250$ km s$^{-1}$ in Figure \ref{fig:feature value} upper panels (left panel:$f_\infty=10$, right panel:$f_\infty/B_{r,\rm{cb}}=0.7$) can be reduced to about $\pm 100$ km s$^{-1}$ in lower panels. The performance of the model should be evaluated through comparisons with observational data, which are left for future work. 

\section{Summary \& Conclusion}

In this study, we conducted one-dimensional simulations of the solar wind from the photosphere to the interplanetary space, and examined the relationship between the flux tube geometry and the solar wind speed at $r = 70R_\odot$. The main findings of this study are summarized as follows:
\begin{enumerate}
\item The simulated solar wind speeds negatively correlate with the asymptotic expansion factor $f_\infty$, as shown in the left panel of Figure \ref{fig:feature value}. This is qualitatively consistent with the Wang-Sheeley empirical model \citep{Wang_1990_ApJ}. However, using only $f_\infty$ to characterize the solar wind speed leads to significant errors, likely because both the radial profile of the flux tube expansion and the magnetic field strength at the coronal base also play key roles. 
\item As shown in Figure \ref{fig:exph_param} and \ref{fig:nonmono_param}, changing the expansion height or scale and non-monotonic expansion result in the significant variation in the solar wind speed. These results are consistent with observational studies such as \citet{Dakeyo_2024_aap, Wang_2019_ApJ}. This indicates that overall $f(r)$ is another important factor in determining the solar wind speed. 
\item Among the feature parameters for solar wind speed, those reflecting the overall profile of the coronal expansion and magnetic field strength show high Pearson correlation coeficients (PCCs). In particular, parameters or integral-based quantities that characterize the expansion height and non-monotonic expansion behavior are suggested to be potentially important for understanding solar wind acceleration. However, it should be noted that these PCCs are derived with equal weighting, and thus the sample distribution differs from that of the actual solar wind. 
\end{enumerate}

Through one-dimensional magnetic flux tube simulations, this study has shown that the solar wind speed correlates strongly with the feature parameters which capture the overall profile of the open flux tube. These parameters can be directly calculated from the radial distribution of the magnetic field $B(r)$ and may contribute to the real-time prediction of solar wind speed in space weather forecasting. 

\begin{acknowledgments}
This work was performed using the CIDAS-computer system of the Institute for Space-Earth Environmental Research, Nagoya University and Cray XC50 and XD2000 at the Center for Computational Astrophysics, National Astronomical Observatory of Japan. KT is supported by International Graduate Program for Excellence in Earth-Space Science (IGPEES), a World-leading Innovative Graduate Study (WINGS) Program, the University of Tokyo. MS and SI are supported by JSPS KAKENHI Grant Numbers JP24K00688, JP25K00976, JP25K01052 and by the grant of Joint Research by the National Institutes of Natural Sciences (NINS) (NINS program No. OML032402). Moreover, this work is supported by JST Sogyo Program (Stage 2), Japan Grant Number JPMJSF2503.
\end{acknowledgments}

\appendix
\section{Effects of smaller $B_{r,\odot}$ }
\label{app:B_rodot}

In the main text, simulations were performed with the photospheric magnetic field strength $B_{r,\odot}$ fixed at $1340$ G. However, weaker magnetic concentrations have been observed on the photosphere \citep{Ishikawa_2021_SciA}, and the actual field strength at the solar wind footpoints likely spans a broader range. This appendix presents additional simulations in which $B_{r,\odot}$ is varied. We carried out calculations for $B_{r,\odot} = 335$ G and $162$ G, considering variations in chromospheric or coronal expansion.

\begin{figure}[!t]
  \begin{center}
  \plotone{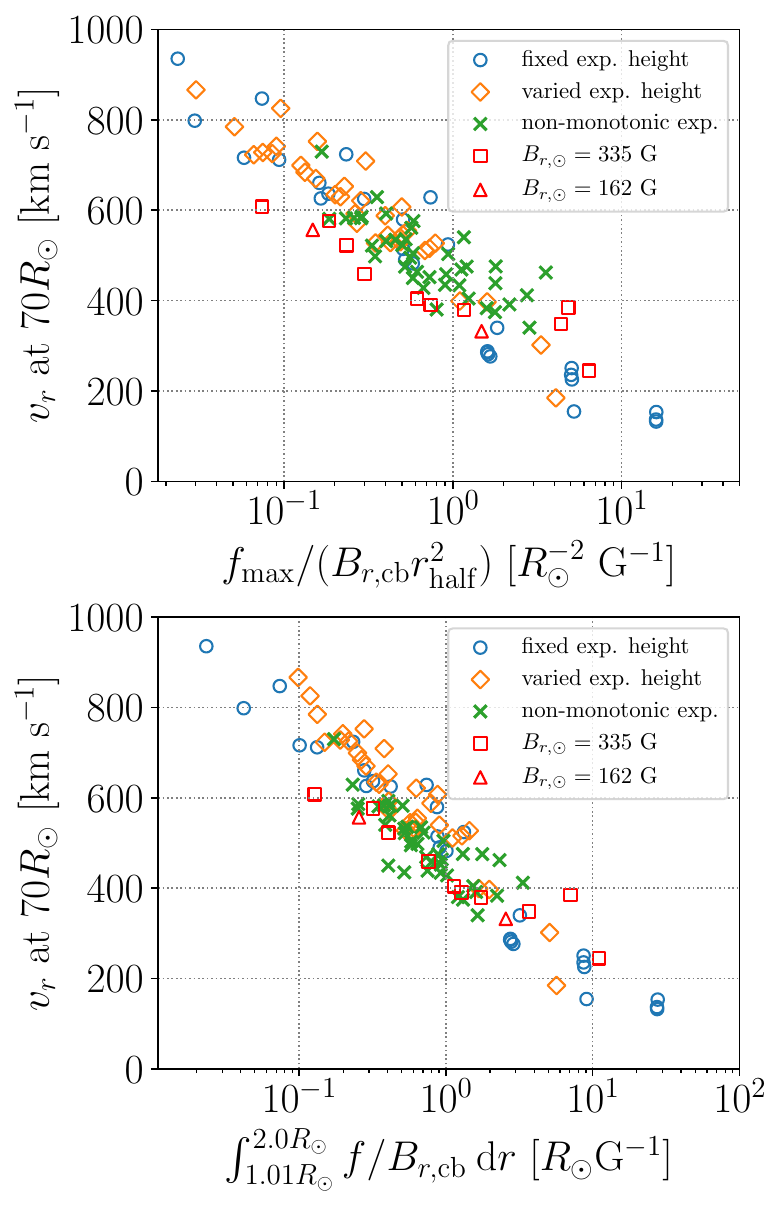}
  \caption{The same as Figure \ref{fig:feature value} but with the results for $B_{r,\odot} = 335$ G and $162$ G to.
  }
  \label{fig:feature value bph}
  \end{center}
\end{figure}
 
Figure \ref{fig:feature value bph} shows the results, including the case with $B_{r,\odot} = 1340$ G. The correlation between the solar wind speed and the quantity $\int_{1.01R_\odot}^{2.0R_\odot}f/B_{r,\rm{cb}}\dd{r}$ remains stable even as $B_{r,\odot}$ varies. Likewise, $f_{\rm{max}}/(B_{r,\rm{cb}}r_{\rm{half}}^2)$ maintains a relatively strong correlation. For $v_r > 400$ km s$^{-1}$, however, lower $B_{r,\odot}$ tends to reduce the solar wind speed, diverging from the trend at $B_{r,\odot} = 1340$ G. As these quantities reflect only coronal magnetic field structures from models such as PFSS, they do not directly account for photospheric variability. This should be kept in mind when applying the approach to observational data.

In varying the photospheric magnetic field, we assumed that the net Alfvén energy flux at the photosphere, $F_{A,\odot,\rm{net}}$, scales linearly with $B_{r,\odot}$, i.e., $F_{A,\odot,\rm{net}} \propto B_{r,\odot}$, implying that $v_{\perp,\odot}$ is independent of $B_{r,\odot}$. However, weaker $B_{r,\odot}$ may induce stronger perturbations $v_{\perp,\odot}$, introducing a possible source of uncertainty in this comparison.
 
\section{Evaluation of a modified $r_{\rm half}$-related feature}
\label{app:r_half}

In the main text, we use the parameter $r_{\rm{half}}^{-2}$ as a representative measure of the expansion height and scale. However, this choice is not unique, and alternative parameterizations should also be considered. In this section, we examine the effects of varying both the definition of expansion height and scale, and the exponent used in the corresponding characteristic parameter.

\begin{figure*}[!t]
  \begin{center}
  \includegraphics[width=18cm]{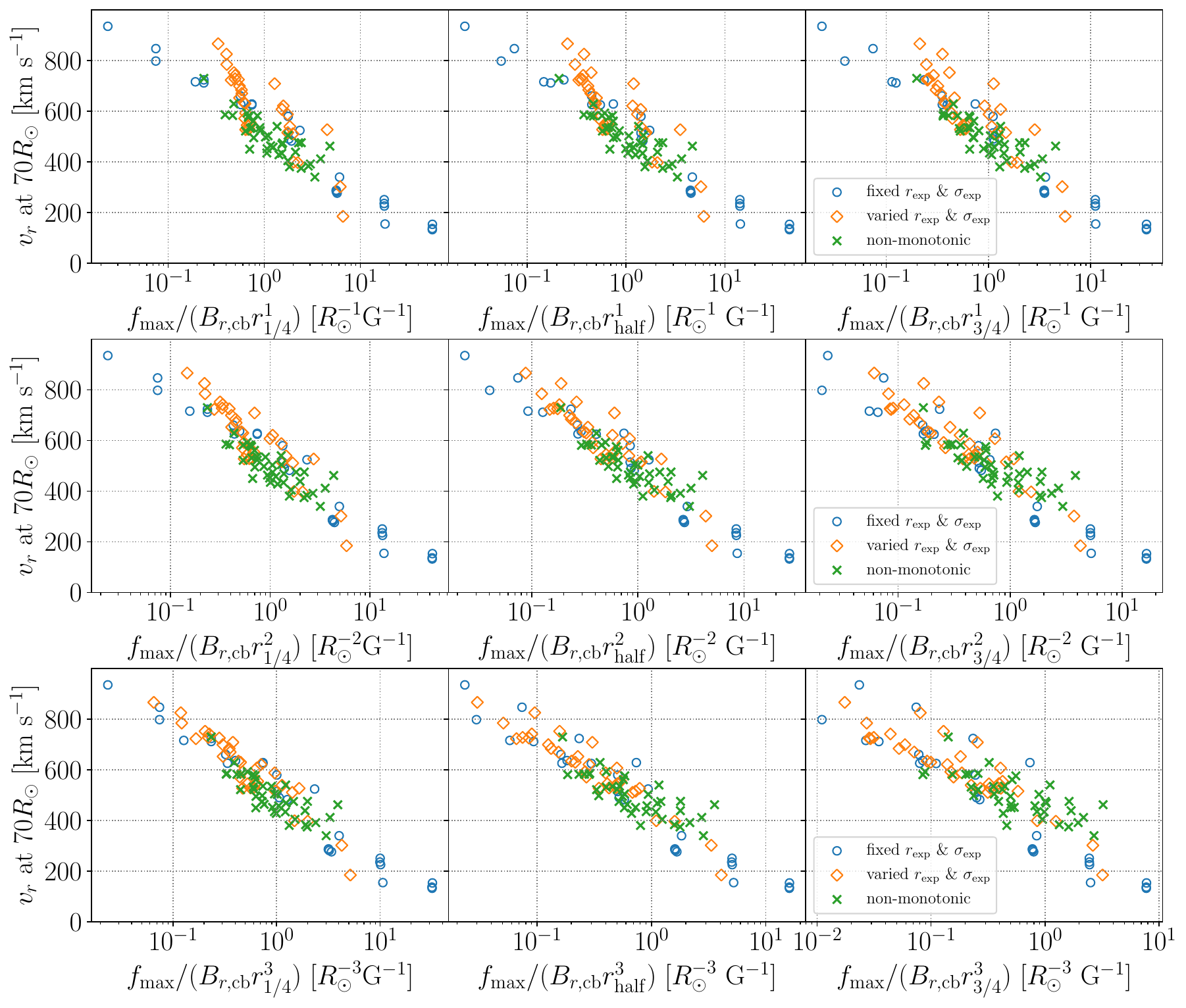}
  \caption{The same as Figure \ref{fig:feature value} but the feature values are limited to $f_{\rm{max}}/(B_{r,\rm{cb}}r_{\rm{X}}^k)$, where X$=1/4,\ 1/2,\ 3/4$ and $k=1,2,3$. $r_{X}$ is the radial distance at which $\left. f(r) \right|_{r=r_{X}}=X f_{\rm{max}}$ for the first time means.}
  \label{fig:r half}
  \end{center}
\end{figure*}

Figure \ref{fig:r half} shows the correlation with the solar wind speed for various definitions of the expansion height: the radial positions where $f(r)$ first reaches $f_{\rm{max}}$ multiplied by $1/4$, $1/2$, and $3/4$, denoted as $r_{1/4}$, $r_{\rm half}$, and $r_{3/4}$, respectively. For each of these, we evaluate characteristic quantities using their $-1$, $-2$, and $-3$ powers. Overall, a higher absolute exponent emphasizes the contribution of the expansion height and scale more strongly, whereas the $-1$ power under-evaluates their influence. The relations between wind speed and $r_{3/4}$-related variables show substantial scatter, likely because the characteristic radius lies too far from the solar surface, making the parameters overly sensitive to small variations in the gradual convergence region. Excluding these cases, the parameters generally correlate well with the solar wind speed.

\section{dependence on the integration region}
\label{app:integration region}

\begin{figure*}[!t]
  \begin{center}
  \plotone{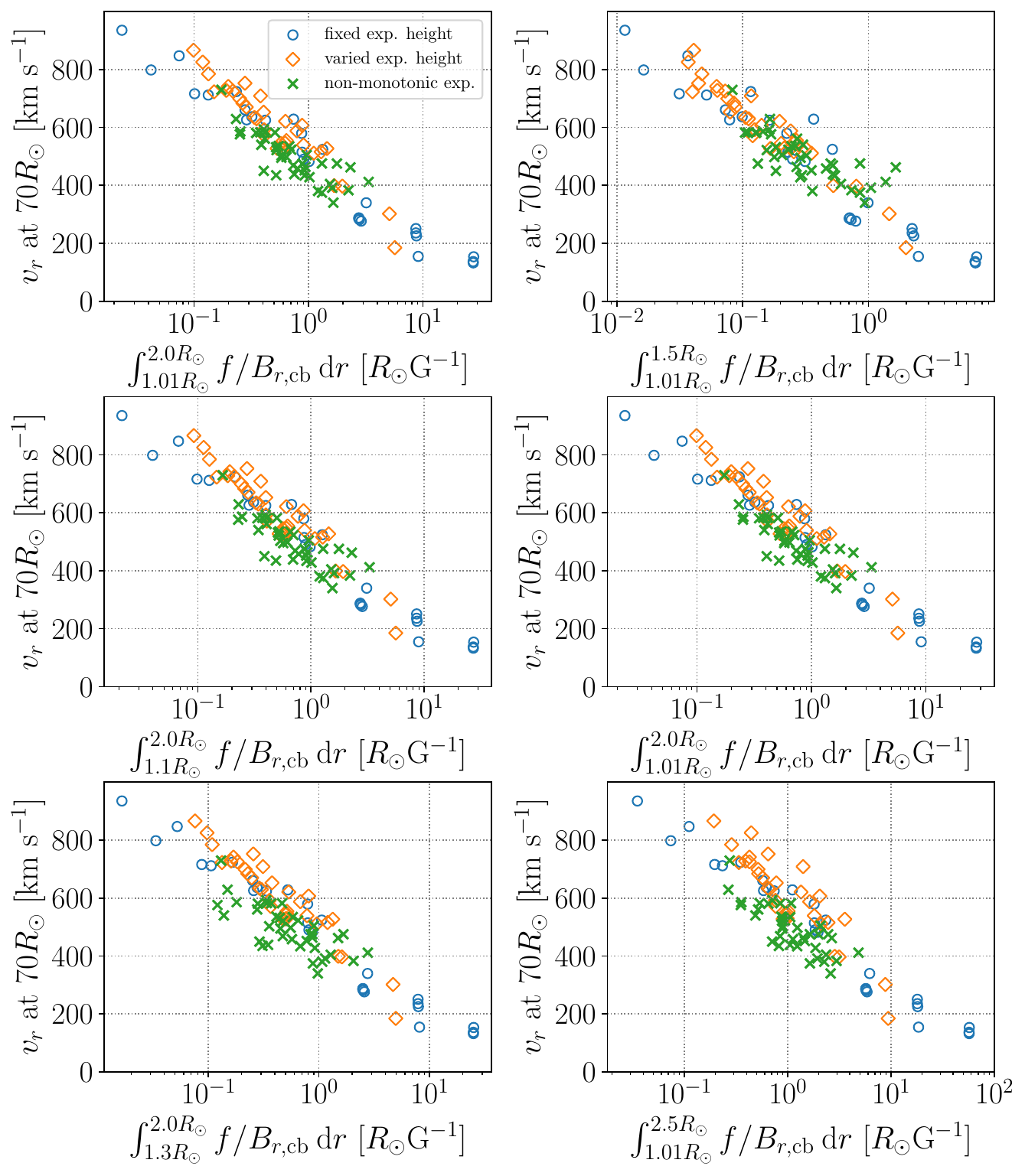}
  \caption{The same as Figure \ref{fig:feature value} but the feature values are limited to $\int_{r_{\rm{bot}}}^{r_{\rm{top}}}f(r)/B_{r,\rm{cb}} \dd{r}$, where $r_{\rm{bot}}=1.01R_\odot,1.1R_\odot,1.3R_\odot$ and $r_{\rm{top}}=1.5R_\odot,2.0R_\odot,2.5R_\odot$. $r_{\rm{bot}}$ varies and $r_{\rm{top}}$ is fixed at $r_{\rm{top}}=2.0R_\odot $in left panels and $r_{\rm{top}}$ varies and $r_{\rm{bot}}$ is fixed at $r_{\rm{bot}}=1.01R_\odot$ in right panels.
  \label{fig:int bot top}}
  \end{center}
\end{figure*}

In the main text, we set the integration range to $1.01R_\odot$–$2.0R_\odot$. However, this choice was made somewhat arbitrarily to achieve strong correlation, and it is necessary to examine the sensitivity of the results to changes in this range. Figure \ref{fig:int bot top} shows the correlation with outer solar wind speed when varying $r_{\rm{bot}}$ (left panel) and $r_{\rm{top}}$ (right panel). In both cases, a certain level of correlation is maintained; however, the overall trend indicates that lower integration heights tend to yield stronger correlations.

\begin{deluxetable}{ccccccccc}
\tablecaption{List of all the simulation parameter settings and the resulting wind velocity at $r=70R_\odot$.
\label{tab:simu_list_1}}
\tablehead{
\colhead{$B_r,\odot$ [G]} & \colhead{ $\eta_{\rm{exp}}$ } & \colhead{ $f_\infty$ } & \colhead{ $r_{\rm{exp}}$ [$R_\odot$] } & \colhead{ $\sigma_{\rm{exp}}$ [$R_\odot$] } & \colhead{ $g_{\rm{max}}$ } & \colhead{ $r_{\rm{max}}$ [$R_\odot$] } & \colhead{ $w$ } & \colhead{ $v_r$ at $70R_\odot$ [km s$^{-1}$]}
}
\startdata
1340.0 & 1.0e+01 & 1.0e+03 & 1.3 & 0.5 & 1.0 & - & - & 284.9 \\ \hline
1340.0 & 1.0e+02 & 1.0e+03 & 1.3 & 0.5 & 1.0 & - & - & 136.4 \\ \hline
1340.0 & 3.2e+01 & 1.0e+03 & 1.3 & 0.5 & 1.0 & - & - & 249.9 \\ \hline
1340.0 & 1.0e+02 & 1.0e+02 & 1.3 & 0.5 & 1.0 & - & - & 281.4 \\ \hline
1340.0 & 1.0e+02 & 1.0e+02 & 1.2 & 0.1 & 1.0 & - & - & 193.6 \\ \hline
1340.0 & 1.0e+02 & 1.0e+02 & 1.3 & 0.1 & 1.0 & - & - & 301.1 \\ \hline
1340.0 & 1.0e+02 & 1.0e+02 & 2.0 & 0.5 & 1.0 & - & - & 525.9 \\ \hline
1340.0 & 1.0e+03 & 1.0e+02 & 1.3 & 0.5 & 1.0 & - & - & 154.7 \\ \hline
1340.0 & 3.2e+01 & 1.0e+02 & 1.3 & 0.5 & 1.0 & - & - & 513.9 \\ \hline
1340.0 & 3.2e+02 & 1.0e+02 & 1.3 & 0.5 & 1.0 & - & - & 224.7 \\ \hline
1340.0 & 1.0e+02 & 1.0e+01 & 1.3 & 0.5 & 1.0 & - & - & 636.1 \\ \hline
1340.0 & 1.0e+02 & 1.0e+01 & 1.05 & 0.1 & 1.0 & - & - & 552.8 \\ \hline
1340.0 & 1.0e+02 & 1.0e+01 & 1.1 & 0.1 & 1.0 & - & - & 544.5 \\ \hline
1340.0 & 1.0e+02 & 1.0e+01 & 1.1 & 0.5 & 1.0 & - & - & 628.9 \\ \hline
1340.0 & 1.0e+02 & 1.0e+01 & 1.2 & 0.1 & 1.0 & - & - & 527.1 \\ \hline
1340.0 & 1.0e+02 & 1.0e+01 & 1.2 & 0.5 & 1.0 & - & - & 633.3 \\ \hline
1340.0 & 1.0e+02 & 1.0e+01 & 1.3 & 0.1 & 1.0 & - & - & 526.0 \\ \hline
1340.0 & 1.0e+02 & 1.0e+01 & 1.3 & 0.3 & 1.0 & - & - & 591.2 \\ \hline
1340.0 & 1.0e+02 & 1.0e+01 & 1.3 & 0.7 & 1.0 & - & - & 683.4 \\ \hline
1340.0 & 1.0e+02 & 1.0e+01 & 1.3 & 1.0 & 1.0 & - & - & 724.3 \\ \hline
1340.0 & 1.0e+02 & 1.0e+01 & 1.5 & 0.1 & 1.0 & - & - & 650.8 \\ \hline
1340.0 & 1.0e+02 & 1.0e+01 & 1.5 & 0.5 & 1.0 & - & - & 669.1 \\ \hline
1340.0 & 1.0e+02 & 1.0e+01 & 1.7 & 0.1 & 1.0 & - & - & 750.2 \\ \hline
1340.0 & 1.0e+02 & 1.0e+01 & 1.7 & 0.5 & 1.0 & - & - & 698.1 \\ \hline
1340.0 & 1.0e+02 & 1.0e+01 & 2.0 & 0.1 & 1.0 & - & - & 824.4 \\ \hline
1340.0 & 1.0e+02 & 1.0e+01 & 2.0 & 0.5 & 1.0 & - & - & 739.8 \\ \hline
1340.0 & 1.0e+02 & 1.0e+01 & 2.0 & 0.7 & 1.0 & - & - & 727.2 \\ \hline
1340.0 & 1.0e+02 & 1.0e+01 & 2.5 & 0.5 & 1.0 & - & - & 784.2 \\ \hline
1340.0 & 1.0e+02 & 1.0e+01 & 3.0 & 0.5 & 1.0 & - & - & 864.2 \\ \hline
1340.0 & 1.0e+03 & 1.0e+01 & 1.3 & 0.5 & 1.0 & - & - & 338.4 \\ \hline
1340.0 & 3.2e+01 & 1.0e+01 & 1.3 & 0.5 & 1.0 & - & - & 715.5 \\ \hline
1340.0 & 3.2e+02 & 1.0e+01 & 1.3 & 0.5 & 1.0 & - & - & 481.2 \\ \hline
1340.0 & 3.2e+02 & 1.0e+01 & 1.1 & 0.1 & 1.0 & - & - & 396.1 \\ \hline
1340.0 & 3.2e+02 & 1.0e+01 & 1.1 & 0.5 & 1.0 & - & - & 509.8 \\ \hline
1340.0 & 3.2e+02 & 1.0e+01 & 1.3 & 0.1 & 1.0 & - & - & 397.4 \\ \hline
1340.0 & 3.2e+02 & 1.0e+01 & 1.5 & 0.1 & 1.0 & - & - & 513.2 \\ \hline
1340.0 & 3.2e+02 & 1.0e+01 & 1.5 & 0.5 & 1.0 & - & - & 539.4 \\ \hline
1340.0 & 3.2e+02 & 1.0e+01 & 1.7 & 0.1 & 1.0 & - & - & 606.4 \\ \hline
1340.0 & 3.2e+02 & 1.0e+01 & 1.7 & 0.5 & 1.0 & - & - & 586.7 \\ \hline
1340.0 & 3.2e+02 & 1.0e+01 & 2.0 & 0.1 & 1.0 & - & - & 708.0 \\ \hline
1340.0 & 3.2e+02 & 1.0e+01 & 2.0 & 0.5 & 1.0 & - & - & 619.2 \\ \hline
1340.0 & 1.0e+02 & 1.0e+00 & 1.3 & 0.5 & 1.0 & - & - & 845.6 \\ \hline
1340.0 & 1.0e+03 & 1.0e+00 & 1.3 & 0.5 & 1.0 & - & - & 628.2 \\ \hline
1340.0 & 3.2e+01 & 1.0e+00 & 1.3 & 0.5 & 1.0 & - & - & 932.2 \\ \hline
1340.0 & 3.2e+02 & 1.0e+00 & 1.3 & 0.5 & 1.0 & - & - & 722.5 \\ \hline
1340.0 & 1.0e+02 & 3.2e+02 & 1.3 & 0.5 & 1.0 & - & - & 235.5 \\ \hline
1340.0 & 3.2e+01 & 3.2e+02 & 1.3 & 0.5 & 1.0 & - & - & 287.1 \\ \hline
1340.0 & 3.2e+02 & 3.2e+02 & 1.3 & 0.5 & 1.0 & - & - & 132.4 \\ \hline
\enddata
\end{deluxetable}

\begin{deluxetable}{ccccccccc}
\tablehead{
\colhead{$B_r,\odot$ [G]} & \colhead{ $\eta_{\rm{exp}}$ } & \colhead{ $f_\infty$ } & \colhead{ $r_{\rm{exp}}$ [$R_\odot$] } & \colhead{ $\sigma_{\rm{exp}}$ [$R_\odot$] } & \colhead{ $g_{\rm{max}}$ } & \colhead{ $r_{\rm{max}}$ [$R_\odot$] } & \colhead{ $w$ } & \colhead{ $v_r$ at $70R_\odot$ [km s$^{-1}$]}
}
\startdata
1340.0 & 1.0e+02 & 3.2e+01 & 1.3 & 0.5 & 1.0 & - & - & 488.5 \\ \hline
1340.0 & 1.0e+03 & 3.2e+01 & 1.3 & 0.5 & 1.0 & - & - & 154.0 \\ \hline
1340.0 & 3.2e+01 & 3.2e+01 & 1.3 & 0.5 & 1.0 & - & - & 625.3 \\ \hline
1340.0 & 3.2e+01 & 3.2e+01 & 1.1 & 0.1 & 1.0 & - & - & 529.2 \\ \hline
1340.0 & 3.2e+01 & 3.2e+01 & 1.2 & 0.1 & 1.0 & - & - & 542.8 \\ \hline
1340.0 & 3.2e+01 & 3.2e+01 & 1.2 & 0.3 & 1.0 & - & - & 570.0 \\ \hline
1340.0 & 3.2e+01 & 3.2e+01 & 2.0 & 0.7 & 1.0 & - & - & 722.1 \\ \hline
1340.0 & 3.2e+02 & 3.2e+01 & 1.3 & 0.5 & 1.0 & - & - & 276.5 \\ \hline
1340.0 & 1.0e+02 & 3.2e+00 & 1.3 & 0.5 & 1.0 & - & - & 710.1 \\ \hline
1340.0 & 1.0e+03 & 3.2e+00 & 1.3 & 0.5 & 1.0 & - & - & 523.8 \\ \hline
1340.0 & 3.2e+01 & 3.2e+00 & 1.3 & 0.5 & 1.0 & - & - & 797.0 \\ \hline
1340.0 & 3.2e+02 & 3.2e+00 & 1.3 & 0.5 & 1.0 & - & - & 624.8 \\ \hline
1340.0 & 1.0e+02 & 3.2e+00 & 1.3 & 0.5 & 16.0 & 1.125 & 2.0 & 536.4 \\ \hline
1340.0 & 1.0e+02 & 3.2e+00 & 1.3 & 0.5 & 16.0 & 1.25 & 2.0 & 466.9 \\ \hline
1340.0 & 3.2e+01 & 3.2e+00 & 1.3 & 0.5 & 16.0 & 1.25 & 2.0 & 626.2 \\ \hline
1340.0 & 3.2e+02 & 3.2e+00 & 1.3 & 0.5 & 16.0 & 1.25 & 2.0 & 461.0 \\ \hline
1340.0 & 1.0e+02 & 3.2e+00 & 1.3 & 0.5 & 2.0 & 1.25 & 2.0 & 728.5 \\ \hline
1340.0 & 3.2e+02 & 3.2e+00 & 1.3 & 0.5 & 2.0 & 1.25 & 2.0 & 535.2 \\ \hline
1340.0 & 3.2e+02 & 3.2e+00 & 1.3 & 0.5 & 2.0 & 1.5 & 2.0 & 533.9 \\ \hline
1340.0 & 3.2e+02 & 3.2e+00 & 1.3 & 0.5 & 2.0 & 2.0 & 2.0 & 497.4 \\ \hline
1340.0 & 3.2e+02 & 3.2e+00 & 1.3 & 0.5 & 4.0 & 1.125 & 2.0 & 502.7 \\ \hline
1340.0 & 1.0e+02 & 3.2e+00 & 1.3 & 0.5 & 4.0 & 1.25 & 2.0 & 584.8 \\ \hline
1340.0 & 3.2e+02 & 3.2e+00 & 1.3 & 0.5 & 4.0 & 1.25 & 2.0 & 457.0 \\ \hline
1340.0 & 1.0e+02 & 3.2e+00 & 1.3 & 0.5 & 4.0 & 1.5 & 2.0 & 581.4 \\ \hline
1340.0 & 3.2e+02 & 3.2e+00 & 1.3 & 0.5 & 8.0 & 1.25 & 2.0 & 373.4 \\ \hline
1340.0 & 1.0e+02 & 3.2e+00 & 1.3 & 0.5 & 8.0 & 1.125 & 2.0 & 574.7 \\ \hline
1340.0 & 1.0e+02 & 3.2e+00 & 1.3 & 0.5 & 8.0 & 1.25 & 2.0 & 560.2 \\ \hline
1340.0 & 3.2e+02 & 3.2e+00 & 1.3 & 0.5 & 8.0 & 1.25 & 2.0 & 474.5 \\ \hline
1340.0 & 1.0e+02 & 3.2e+00 & 1.3 & 0.5 & 8.0 & 1.5 & 2.0 & 521.3 \\ \hline
1340.0 & 3.2e+02 & 3.2e+00 & 1.3 & 0.5 & 8.0 & 1.5 & 2.0 & 383.0 \\ \hline
1340.0 & 1.0e+02 & 1.0e+01 & 1.3 & 0.5 & 1.5 & 2.0 & 2.0 & 581.0 \\ \hline
1340.0 & 1.0e+02 & 1.0e+01 & 1.3 & 0.5 & 16.0 & 1.125 & 2.0 & 439.6 \\ \hline
1340.0 & 1.0e+02 & 1.0e+01 & 1.3 & 0.5 & 16.0 & 1.25 & 2.0 & 336.9 \\ \hline
1340.0 & 1.0e+02 & 1.0e+01 & 1.3 & 0.5 & 16.0 & 1.25 & 2.0 & 357.6 \\ \hline
1340.0 & 1.0e+02 & 1.0e+01 & 1.3 & 0.5 & 16.0 & 1.25 & 2.0 & 391.0 \\ \hline
1340.0 & 1.0e+02 & 1.0e+01 & 1.3 & 0.5 & 16.0 & 1.25 & 2.0 & 378.5 \\ \hline
1340.0 & 1.0e+02 & 1.0e+01 & 1.3 & 0.5 & 16.0 & 1.25 & 2.0 & 359.9 \\ \hline
1340.0 & 1.0e+02 & 1.0e+01 & 1.3 & 0.5 & 2.0 & 1.125 & 2.0 & 580.2 \\ \hline
1340.0 & 1.0e+02 & 1.0e+01 & 1.3 & 0.5 & 2.0 & 1.25 & 2.0 & 591.3 \\ \hline
1340.0 & 1.0e+02 & 1.0e+01 & 1.3 & 0.5 & 2.0 & 1.5 & 2.0 & 520.3 \\ \hline
1340.0 & 1.0e+02 & 1.0e+01 & 1.3 & 0.5 & 2.0 & 2.0 & 2.0 & 582.5 \\ \hline
1340.0 & 1.0e+02 & 1.0e+01 & 1.3 & 0.5 & 4.0 & 1.125 & 2.0 & 449.1 \\ \hline
1340.0 & 1.0e+02 & 1.0e+01 & 1.3 & 0.5 & 4.0 & 1.25 & 2.0 & 494.3 \\ \hline
1340.0 & 1.0e+02 & 1.0e+01 & 1.3 & 0.5 & 4.0 & 1.5 & 2.0 & 462.6 \\ \hline
1340.0 & 1.0e+02 & 1.0e+01 & 1.3 & 0.5 & 4.0 & 2.0 & 2.0 & 474.2 \\ \hline
1340.0 & 1.0e+02 & 1.0e+01 & 1.3 & 0.5 & 8.0 & 1.125 & 2.0 & 434.1 \\ \hline
1340.0 & 3.2e+02 & 1.0e+01 & 1.3 & 0.5 & 8.0 & 1.125 & 2.0 & 340.1 \\ \hline
1340.0 & 1.0e+02 & 1.0e+01 & 1.3 & 0.5 & 8.0 & 1.25 & 2.0 & 361.7 \\ \hline
1340.0 & 1.0e+02 & 1.0e+01 & 1.3 & 0.5 & 8.0 & 1.25 & 2.0 & 408.7 \\ \hline
\enddata
\end{deluxetable}

\begin{deluxetable}{ccccccccc}
\tablehead{
\colhead{$B_r,\odot$ [G]} & \colhead{ $\eta_{\rm{exp}}$ } & \colhead{ $f_\infty$ } & \colhead{ $r_{\rm{exp}}$ [$R_\odot$] } & \colhead{ $\sigma_{\rm{exp}}$ [$R_\odot$] } & \colhead{ $g_{\rm{max}}$ } & \colhead{ $r_{\rm{max}}$ [$R_\odot$] } & \colhead{ $w$ } & \colhead{ $v_r$ at $70R_\odot$ [km s$^{-1}$]}
}
\startdata
1340.0 & 1.0e+02 & 1.0e+01 & 1.3 & 0.5 & 8.0 & 1.25 & 2.0 & 432.9 \\ \hline
1340.0 & 1.0e+02 & 1.0e+01 & 1.3 & 0.5 & 8.0 & 1.25 & 2.0 & 401.5 \\ \hline
1340.0 & 1.0e+02 & 1.0e+01 & 1.3 & 0.5 & 8.0 & 1.5 & 2.0 & 473.1 \\ \hline
1340.0 & 1.0e+02 & 1.0e+01 & 1.3 & 0.5 & 16.0 & 1.25 & 4.0 & 409.2 \\ \hline
1340.0 & 1.0e+02 & 1.0e+01 & 1.3 & 0.5 & 2.0 & 1.25 & 4.0 & 530.4 \\ \hline
1340.0 & 1.0e+02 & 1.0e+01 & 1.3 & 0.5 & 4.0 & 1.25 & 4.0 & 451.2 \\ \hline
1340.0 & 1.0e+02 & 1.0e+01 & 1.5 & 0.5 & 4.0 & 1.5 & 4.0 & 427.2 \\ \hline
1340.0 & 1.0e+02 & 1.0e+01 & 1.5 & 0.5 & 4.0 & 2.0 & 4.0 & 505.4 \\ \hline
1340.0 & 1.0e+02 & 1.0e+01 & 1.5 & 0.5 & 8.0 & 1.25 & 4.0 & 404.4 \\ \hline
1340.0 & 1.0e+02 & 1.0e+01 & 1.3 & 0.5 & 4.0 & 1.5 & 8.0 & 379.5 \\ \hline
335.0 & 1.0e+02 & 1.0e+02 & 1.3 & 0.5 & 1.0 & - & - & 243.1 \\ \hline
335.0 & 1.0e+01 & 1.0e+01 & 1.3 & 0.5 & 1.0 & - & - & 606.3 \\ \hline
335.0 & 1.0e+02 & 1.0e+01 & 1.3 & 0.5 & 1.0 & - & - & 389.4 \\ \hline
335.0 & 1.0e+02 & 1.0e+01 & 1.2 & 0.3 & 1.0 & - & - & 376.4 \\ \hline
335.0 & 1.0e+02 & 1.0e+01 & 1.5 & 0.5 & 1.0 & - & - & 402.6 \\ \hline
335.0 & 1.0e+02 & 1.0e+01 & 2.0 & 0.7 & 1.0 & - & - & 458.1 \\ \hline
335.0 & 3.2e+01 & 1.0e+01 & 1.3 & 0.5 & 1.0 & - & - & 521.5 \\ \hline
335.0 & 2.5e+01 & 1.0e+01 & 1.3 & 0.5 & 1.0 & - & - & 575.5 \\ \hline
335.0 & 1.0e+02 & 1.0e+01 & 1.3 & 0.5 & 8.0 & 1.25 & 2.0 & 347.3 \\ \hline
335.0 & 1.0e+02 & 1.0e+01 & 1.3 & 0.5 & 8.0 & 1.5 & 2.0 & 383.1 \\ \hline
167.5 & 1.0e+01 & 1.0e+01 & 1.3 & 0.5 & 1.0 & - & - & 555.3 \\ \hline
167.5 & 1.0e+02 & 1.0e+01 & 1.3 & 0.5 & 1.0 & - & - & 332.7 \\ \hline
\enddata
\end{deluxetable}

\clearpage

\bibliography{reference}{}
\bibliographystyle{aasjournal}

\end{document}